\documentclass{article}
\usepackage{etoolbox}
\usepackage{amsmath,amssymb,amsfonts}
\usepackage{algorithmic}
\usepackage{graphicx,color}
\usepackage{textcomp}
\usepackage{xcolor}
\usepackage{hyperref}
\usepackage{comment}
\usepackage{tikz}
\usetikzlibrary{shapes.geometric, arrows}
\tikzstyle{block} = [rectangle, rounded corners, minimum width=3.5cm, minimum height=1cm,text centered, draw=black, fill=blue!10]
\tikzstyle{process} = [rectangle, minimum width=3.5cm, minimum height=1cm,text centered, draw=black, fill=green!10]
\tikzstyle{decision} = [diamond, aspect=2, text centered, draw=black, fill=orange!15]
\tikzstyle{arrow} = [thick,->,>=stealth]

\usepackage{algorithm,algorithmic}
\usepackage{amsthm}

\usepackage[ natbib=true,style=numeric,sorting=none]{biblatex}
\def\BibTeX{{\rm B\kern-.05em{\sc i\kern-.025em b}\kern-.08em
    T\kern-.1667em\lower.7ex\hbox{E}\kern-.125emX}}
\AtBeginDocument{\definecolor{tmlcncolor}{cmyk}{0.93,0.59,0.15,0.02}\definecolor{NavyBlue}{RGB}{0,86,125}}
\usepackage{hyperref}
\hypersetup{
    colorlinks=true, 
    linkcolor=black, 
    citecolor=blue,  
    filecolor=magenta, 
    urlcolor=blue    
}
\definecolor{lime}{HTML}{A6CE39}

\DeclareUnicodeCharacter{2217}{\ast}

\def\OJlogo{\vspace{-4pt}\includegraphics[height=18pt]{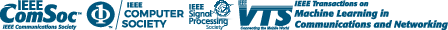}}
\def\seclogo{\vspace{10pt}\includegraphics[height=18pt]{new_logo}}

\begin{document}


\title{A Deep-SIC Channel Estimator Scheme in NOMA Network}

\title{A Deep-SIC Channel Estimator Scheme in NOMA Network}

\author{%
    Sumita Majhi\textsuperscript{1},
    Kaushal Shelke\textsuperscript{2},
    and Pinaki Mitra\textsuperscript{1}
}
\date{%
    \textsuperscript{1}Department of Computer Science and Engineering\\
    Indian Institute of Technology Guwahati\\
    Assam, India\\
    \textsuperscript{2}Department of Computer Science and Engineering\
    Indian Institute of Information Technology, Bhagalpur\\
    Bihar, India \\
    Email: \textsuperscript{1}\{sumita, pinaki\}@iitg.ac.in,
    \textsuperscript{2}kaushal.2101051cs@iiitbh.ac.in
}

\begin{abstract}
In 5G and next-generation mobile ad-hoc networks, reliable handover is a key requirement, which guarantees continuity in connectivity, especially for mobile users and in high-density scenarios. However, conventional handover triggers based on instantaneous channel measurements are prone to failures and the ping-pong effect due to outdated or inaccurate channel state information. To address this, we introduce Deep-SIC, a knowledge-based channel prediction model that employs a Transformer-based approach to predict channel quality and optimise handover decisions. Deep-SIC is a unique model that utilises Partially Decoded Data (PDD), a byproduct of successive interference cancellation (SIC) in NOMA, as a feedback signal to improve its predictions continually. This special purpose enables learners to learn quickly and stabilise their learning. Our model learns 68\% faster than existing state-of-the-art algorithms, such as Graph-NOMA, while offering verifiable guarantees of stability and resilience to user mobility (Theorem~2). When simulated at the system level, it can be shown that our strategy can substantially enhance network performance: the handover failure rate can be reduced by up to 40\%, and the ping-pong effect can be mitigated, especially at vehicular speeds (e.g., 60 km/h). Moreover, Deep-SIC has a 20\% smaller normalised root mean square error (NRMSE) in low-SNR situations than state-of-the-art algorithms with linear computational complexity, $O(K)$. This work has introduced a new paradigm for robust and predictive mobility management in dynamic wireless networks.

\end{abstract}


\maketitle

\begin{table}[ht]
\caption{Abbreviation Table}
\begin{center}
\begin{tabular}{|c|c|}
\hline
 Abbreviation &  Definition \\
 \hline
 NOMA  & Non-Orthogonal Multiple Access \\ 
 \hline
 SIC  & Successive Interference Cancellation \\ 
 \hline
  CSI  & Channel State Information \\ 
 \hline
  PDD  & Partially Decoded Data  \\ 
 \hline
  HOF  & Handover Failure \\ 
  \hline
  RSRQ  & Reference Signal Receive Quality \\ 
 \hline
  MMSE  & Minimum Mean Square Error \\ 
  \hline
  SINR  & Signal-to-Interference-plus-Noise Ratio  \\ 
 \hline
 NOMA-HO  & NOMA Handover  \\ 
 \hline
 UE  & User Equipment  \\ 
 \hline
SC & Superposition Coding \\
\hline
 BS & Base Station \\
\hline
DnCNN & Denoising Convolutional Neural Network \\
 \hline
  SINR  & Signal-to-Interference-plus-Noise Ratio  \\ 
 \hline
 NRMSE  & Normalized Root Mean Squared Error   \\ 
 \hline
MASE & Mean Absolute Scaled Error  \\
\hline
MSE & Mean Squared Error \\
\hline
 $R^{2}$ & R-squared  \\
  \hline
 TTT & Time to Trigger \\
  \hline
 BER & Bit Error Rate \\
   \hline
\end{tabular}
\end{center} 
\end{table}

\section{INTRODUCTION}
The proliferation of mobile devices and the advent of IoT have made seamless connectivity a cornerstone of modern wireless systems. In mobile ad-hoc and cellular networks, user mobility necessitates frequent handovers between base stations or access points. These handovers are not very reliable, however, and cause dropped connections, poor quality of service and the ping-pong effect- where cell to cell switches between users. The primary causes of these failures are the reliance on outdated channel state information (CSI) to make decisions during handovers, which fail to account for the rapid temporal changes resulting from user movement and fading. Conventional handover algorithms are predictively inactive, as they activate based on simple thresholds of metrics such as Reference Signal Received Quality (RSRQ). They respond to channel conditions rather than predicting them.

\par
Traditional handover algorithms, which trigger based on simple thresholds of metrics like Reference Signal Received Quality (RSRQ), lack predictive capability. They react to channel conditions rather than anticipating them. While recent studies have explored machine learning for channel estimation, many focus on denoising or static scenarios. They are not aimed at the long-term forecasting that is needed to hand over proactively. Moreover, the issue is worsened in Non-Orthogonal Multiple Access (NOMA) networks, which is one of the key technologies of the 5G, as the interference of users is complicated, and channel prediction becomes even more difficult but essential.
\par
In order to address these shortcomings, we present Deep-SIC, a novel predictive substructure of intelligent handover management. At its core is a Transformer-based channel predictor, a model that captures long-term temporal relationships to predict future channel quality accurately. In a novel way, Deep-SIC creates a closed feedback system that operates based on Partially Decoded Data (PDD) of the NOMA SIC process. This PDD serves as an implicit error signal, enabling the model to self-correct and enhance the reliability of its predictions in real-time. By doing so, our approach transforms the handover process from a reactive to a predictive and reliable operation. The main contributions of this work are:

\begin{itemize}
\item A predictive handover management framework (Deep-SIC) that uses a Transformer model for accurate short-term channel forecasting in mobile environments.

\item A new closed-loop feedback system with PDD is used to make the prediction more robust and minimise the effects of interference cancellation errors.

\item Theoretical convergence and mobility resilience guarantees of the model ensure stability under dynamic network conditions.

\item Extensive system-level simulations demonstrate that Deep-SIC substantially reduces handover failure rates (by up to 40\%) and ping-pong effects, while improving spectral efficiency which is a key metric for network operators.
\end{itemize}
The remainder of this paper is organized as follows: Section II reviews related work. Section III presents the motivation and contributions. Section IV introduces the system model. Section V describes the Deep-SIC framework. Section VI discusses results, Section VII provides theoretical analysis, and Section VIII concludes the paper.

\section{Related work}
The quality of connectivity in mobile ad-hoc and cellular networks is mainly based on reliable channel estimation, which directly influences such important functions as power control, interference management, and, most importantly, smooth handover. The problem of acquiring valid Channel State Information (CSI) is exacerbated when using dynamic NOMA-based networks, as several users share resources and are typically mobile. The classical estimation methods, although fundamental, display severe restrictions, as we examine below.
\par
Classical estimators such as the Minimum Mean Square Error (MMSE) and Least Squares (LS) remain reference points in the literature \cite{34,35}. MMSE achieves optimality under Gaussian noise when the channel statistics are known, but its implementation often requires large matrix inversions, which become costly in high-dimensional MIMO settings. The LS estimator is computationally simpler and model-agnostic, yet it degrades markedly in low-SNR regimes and is more susceptible to interference and noise.
\par
Much recent work has focused on improving pilot-based estimation for modern deployments (e.g., MIMO, RIS, massive MIMO). For example, Meenalakshmi et al. \cite{27} integrate a CNN with polar coding to lower MSE and BER in MIMO–OFDM; Chen et al. \cite{28} propose an LS-based estimator with tailored training to limit the impact of amplified thermal noise in active RIS; and tensor-decomposition approaches by Gomes et al. \cite{29} (TALS, HOSVD) have shown robustness to specific hardware impairments in RIS-assisted systems. Despite these advances, pilot-based approaches still carry unavoidable overhead, remain sensitive to modelling mismatch, and lose accuracy when the channel varies rapidly, conditions that are typical during user mobility and handover. 
\par

Blind estimation methods avoid pilots by exploiting the statistical structure of the received signals. Representative classes include subspace-based techniques \cite{48,52,53,54} and higher-order-statistics (HOS) methods \cite{44,46,47}. Lawal et al. \cite{45,52} introduce a Structured Signal Subspace (SSS) approach that leverages Toeplitz structure for more robust estimation under ill-conditioned channels, while Gong et al. \cite{52} enhance ESPRIT with tensor-train decompositions for mmWave 3D MIMO-OFDM.The blind methods are often highly computational in nature; e.g., Liu et al. \cite{47} report an $O(N^3)$ complexity due to matrix inversions and hence cannot be applicable in real-time applications, such as resource-constrained and handover scenarios.
\par

Semi-blind methods aim to strike a balance between dense pilot data and data-driven information. Examples are the antenna-partitioned scheme of Alwakeel and Mehana \cite{57} of massive MIMO, reinforcement-learning-based estimators by Jeon et al. \cite{59}, and iterative virtual-pilot schemes in Park et al. \cite{56}. These approaches reduce pilot overhead but introduce a new vulnerability: if early symbol detections are incorrect, using those detections as pseudo-pilots can contaminate the channel estimate and trigger error propagation.
\par
This weakness in propagating errors highlights a significant flaw in most designs based on semi-blindness: such designs lack effective methods for determining which symbols derived from data are reliable. This is the weakness to be overcome, and this inspires our structure. In particular, Deep-SIC utilises the learned trustworthiness metric, which eliminates PDD in the SIC process and removes unreliable symbols that negatively impact the estimate, thereby enhancing consistency and accuracy.
\par
Additionally, machine learning architectures, such as Denoising CNNs (DnCNNs) \cite{32}, also have their benefits. However, the localised receptive field of this type does not mean that the long-range temporal dependencies, which arise due to user mobility, are considered. This renders them inappropriate for predictive forecasting to proactively control the handovers and avoid connection dropouts in a dynamic network.
\par
In the context of NOMA specifically, research addressing NOMA handover processes (NOMA-HO) and the complexity of managing multiple users with varying power levels during a handover remains limited. Several prior studies \cite{8,9,10,12} focus on CSI estimation for NOMA but typically rely on conventional CSI-centric methods that are not designed for the rapid forecasting required by handover control. Practical mobility problems—handover failures (HOF) and ping-pong effects—remain important concerns in NOMA networks \cite{10,18}. They occur when a cell transition of a UE is premature or when channel variability causes transient oscillatory handovers, resulting in lost connections and a poor user experience. Precise prediction of the short-term CSI is thus useful, as this enables the network to perform handovers more intelligently and prevent transitions that have a high likelihood of failure.
\par
\begin{figure}[htbp]
    \centering
    \includegraphics[width=0.5\textwidth]{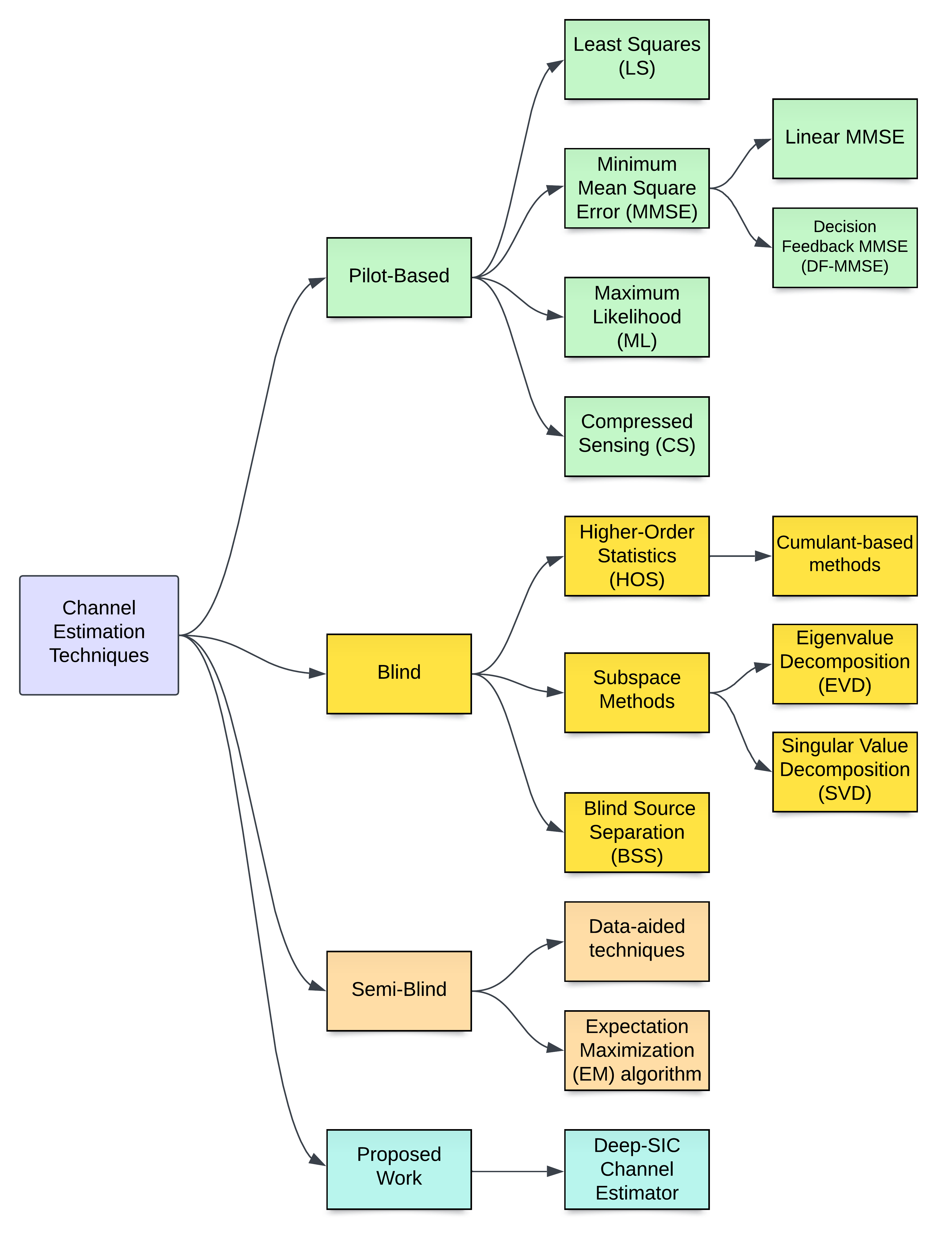} 
    \caption{Classification of Channel Estimation Techniques.}
    \label{1}
\end{figure}

Overall, the current methods, as demonstrated in Figure~\ref{1}, cannot be effectively employed within the constraints of mobile NOMA networks. What is required is an estimator not only precise but also predictive, so that proactive handovers can be made. At the same time, it must be resource-efficient to ensure the network capacity is not exceeded, and resilient to the errors that are bound to occur in a dynamic environment. This gap is especially pronounced for the NOMA handover (NOMA-HO) scenario, which has received limited attention despite its practical importance.
\par
These observations point to a clear set of requirements for a practical estimator in NOMA handover scenarios:
\begin{itemize}
\item Temporally aware: able to model long-range dependencies for accurate short-term forecasting during mobility. 
\item Immune to imperfections: made to take advantage, not to be fooled by, imperfect SIC outputs.
\item Both the convergence and stability of the models are theoretical and valid under realistic operating conditions.
\item Handover-based: adapted to reduce the HOF and ping-pong rates based on anticipated CSI estimations.
\end{itemize}
To the best of our knowledge, there is no prior work that combines a Transformer-style temporal model with a PDD-based feedback loop derived from the SIC process to produce a closed-loop, self-correcting channel estimator tailored for NOMA handovers. The Deep-SIC framework proposed in this paper aims to close that gap by uniting temporal representation learning, data-aided reliability assessment, and theoretical performance bounds.

\section{Motivation and contributions}
Handover failures (HOF) and the ping-pong effect are primary causes of dropped connections and poor user experience in mobile networks. These events are often a direct consequence of outdated channel state information, which prevents the network from making optimal handover decisions. While deep learning models like DnCNN \cite{32} can mitigate static impairments, they are unable to model the long-term temporal evolution of the channel caused by user mobility—precisely the information needed to predict and prevent a faulty handover.
\par
The requirement for a temporally aware and imperfection-adaptive channel estimator is what motivated this work. The Transformer architecture naturally fits this goal: Its self-attention mechanism lets the model weigh past channel states and pick out the most relevant history to predict what comes next. That capability makes it well-suited for short-term forecasting of channel behavior, which is what a handover controller needs to act ahead of time.
\par
In addition, we challenge a common assumption: residual interference left after imperfect SIC is usually treated as noise and ignored. In practice, the partially decoded symbols (PDD) contain a feedback signal about the error in the initial channel estimate. In other words, PDD becomes feedback that helps the estimator notice and correct its mistakes.
\par
Thus, our work pursues two linked goals. First, we replace purely local models with a Transformer-based estimator that captures long-range temporal dependencies and produces accurate, short-term CSI predictions. Second, we close the loop between detection and estimation by turning PDD into a learned feedback signal that continuously refines the channel estimate and reduces sensitivity to SIC errors and mobility.
\par
To the best of our knowledge, prior work \cite{10,18,calhan2020handover,abhirami2024handover,marcano2018impact,liu2022evolution} has not combined a Transformer with PDD-based feedback for NOMA handover. Our contributions are listed below.
\begin{itemize}
    \item We propose Deep-SIC, a predictive channel estimation framework (Algorithm 1) that reduces handover failure (HOF) rates by up to 40\% and mitigates the ping-pong effect in mobile NOMA networks.
    \item The key enabler is a novel closed-loop feedback system that leverages Partially Decoded Data (PDD) from the SIC process to continuously correct and refine channel predictions.
    \item A Transformer-based temporal model, which is at the centre of the framework, is an effective method of predicting channel conditions, including the consideration of long-range dependencies, and, as a result, can make a decision concerning the network proactively.
    \item We come up with theoretical guarantees of our model: convergence conditions (Proposition 1), estimation error bounds (Theorem 1), a mobility resilience bound (Theorem 2), and stability and predictability in dynamic environments.

    \item We demonstrate that our approach is practically feasible, scaling with linear complexity $O(K)$ in the number of users (Proposition 2, Fig. 12), and leverage transfer learning to ensure robust performance even with limited data.
\end{itemize}

\section{System Model}
\label{sec:system_model}
We consider a downlink NOMA network scenario representative of a mobile cellular environment, where User Equipments (UEs) must maintain connectivity while moving. A single base station (BS) equipped with \(M\) antennas serves \(K\) single-antenna UEs. The channel from the BS to the \(k\)-th UE is a complex vector \(\mathbf{h}_k \in \mathbb{C}^{M\times 1}\)  , which we model as Rayleigh fading to capture the signal variations typical in mobile scenarios, i.e., \(\mathbf{h}_k \sim \mathcal{CN}(\mathbf{0},\mathbf{I}_M)\). The BS sends a superposed signal
\[
\mathbf{x} = \sum_{k=1}^{K} \sqrt{P_k}\,\mathbf{w}_k s_k,
\]
where \(s_k\) denotes the data symbol, \(P_k\) the transmit power, and \(\mathbf{w}_k\in\mathbb{C}^{M\times 1}\) the unit-norm precoding vector for user \(k\) (\(\|\mathbf{w}_k\|^2 = 1\)). The received signal at user \(k\) is
\begin{equation}
y_k = \mathbf{h}_k^{H}\mathbf{x} + n_k,
\label{eq:received_signal}
\end{equation}
or equivalently
\begin{equation}
y_k = \mathbf{h}_k^{H}\Bigg(\sum_{i=1}^{K}\sqrt{P_i}\,\mathbf{w}_i s_i\Bigg) + n_k,
\label{eq:received_signal_expanded}
\end{equation}
with \(n_k \sim \mathcal{CN}(0,\sigma_n^2)\) modeling additive white Gaussian noise (AWGN).

We order users by their effective channel gains:
\[
\big|\mathbf{h}_1^{H}\mathbf{w}_1\big|^2 \ge \big|\mathbf{h}_2^{H}\mathbf{w}_2\big|^2 \ge \cdots \ge \big|\mathbf{h}_K^{H}\mathbf{w}_K\big|^2.
\]
Following the NOMA protocol, the strongest user (User 1) performs SIC. User 1 first decodes the signals intended for weaker users (Users 2 through \(K\)), subtracts those estimates from its received signal, and then decodes its own symbol. A weaker user \(k>1\) treats signals from stronger users as interference and decodes its symbol directly.
\begin{figure}[htbp]
    \centering
    \includegraphics[width=0.5\textwidth]{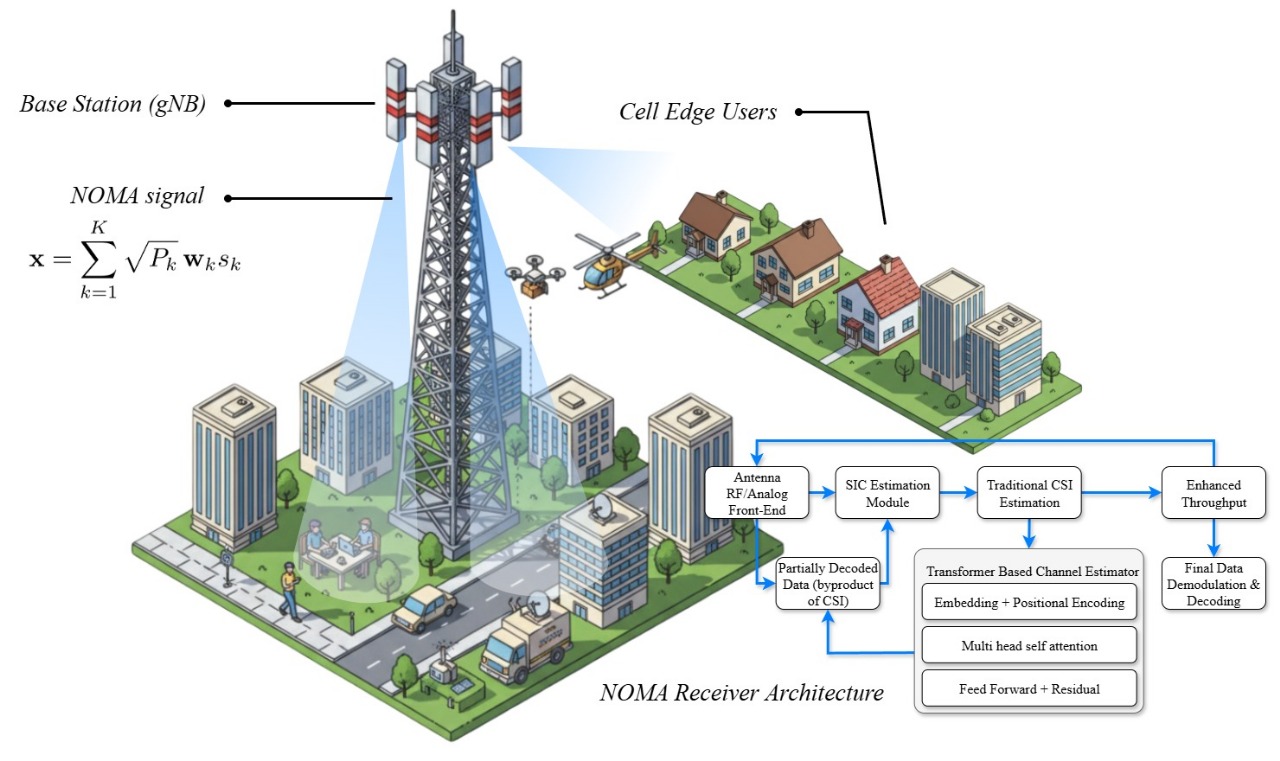} 
    \caption{System Model and Receiver Architecture.}
    \label{1_222}
\end{figure}
\subsection{Imperfect SIC and Partially Decoded Data}
SIC relies on good channel estimates \(\hat{\mathbf{h}}_k\). When these estimates are imperfect, cancellation leaves residual interference. To capture this effect, we introduce PDD, a byproduct of CSI in NOMA channel estimation. Let \(\tilde{s}_i^{(k)}\) be the symbol for user \(i\) as decoded by user \(k\) during SIC. For example, when User 1 attempts to decode User 2, the decoding error is
\[
\widetilde{s}_2 = s_2 - \tilde{s}_2^{(1)},
\]
so \(\widetilde{s}_2\) denotes the residual, undecoded component (the PDD). This residual interference not only degrades the user's immediate signal decoding but also corrupts the channel quality estimates used for long-term network management tasks like handover. Therefore, we propose to use this PDD as a feedback signal to correct the initial channel estimate, thereby improving the overall stability of the network's view of the channel. After attempting to cancel User 2's contribution, the received signal at User 1 becomes
\begin{equation}
y_{1,\text{SIC}} = \mathbf{h}_1^{H}\big(\sqrt{P_1}\,\mathbf{w}_1 s_1 + \sqrt{P_2}\,\mathbf{w}_2 \widetilde{s}_2\big)
+ \sum_{i=3}^{K}\mathbf{h}_1^{H}\sqrt{P_i}\,\mathbf{w}_i s_i + n_1.
\label{eq:strong_user_sic}
\end{equation}
The term \(\sqrt{P_2}\,\mathbf{w}_2 \widetilde{s}_2\) is the residual interference that degrades User 1's decoding. Rather than treating this term as pure noise, we view \(\widetilde{s}_2\) as a useful signal that carries information about the error in the channel estimate. We use this PDD as an additional feedback measure to improve the initial channel estimate.

For a weaker user (e.g., User 2), the received signal is
\begin{equation}
y_2 = \mathbf{h}_2^{H}\sqrt{P_1}\,\mathbf{w}_1 s_1 + \mathbf{h}_2^{H}\sqrt{P_2}\,\mathbf{w}_2 s_2
+ \sum_{i=3}^{K}\mathbf{h}_2^{H}\sqrt{P_i}\,\mathbf{w}_i s_i + n_2,
\label{eq:weak_user}
\end{equation}
and User 2 must treat the first term as interference since it does not perform SIC.

\subsection{NOMA Handover (NOMA-HO) Scenario}
Robust channel estimates are especially important during handover. We consider a two-cell setup where a far user \(U_0^F\) currently served by \(BS_0\) may hand over to a neighboring cell served by \(BS_1\). A handover decision typically compares channel-quality metrics such as Reference Signal Received Quality (RSRQ) derived from CSI. We improve this choice by considering the stability of each estimate using PDD. The efficiency of this system is assessed in a mobility situation where a user is required to decide to pass on the serving BS to a target BS, where the accuracy and predictability of the channel estimates $h_{k}$ hold the success of the process. In practice, we cause a handover when the adjusted quality of the target BS is better than the quality of the serving BS:
\begin{equation}
\text{CSI}_1 + \alpha\cdot\text{PDD}_1 \;>\; \text{CSI}_0 + \alpha\cdot\text{PDD}_0,
\label{eq:ho_decision}
\end{equation}
where \(\text{CSI}_j\) represents the estimated quality of all the channels of BS \(j\), \(\text{PDD}_j\) represents the measurement of estimation reliability (a smaller value signifies a more reliable estimate), and \(\alpha\) is a scaling factor to reduce compromises between quality and reliability. Under this rule, the network assigns users to cells with stronger signals, as well as to cells whose channel estimations are more reliable, thereby reducing ping-pong effects and failures during handover.

\section{Proposed Work}
The Deep-SIC framework is an intelligent processing module in the network stack that supports handover decisions. It operates in two primary stages: initially, it performs fundamental channel estimation using regular pilots. Then it enters a refinement loop, where it utilises the data transmission itself to refine the estimation. It is this self-correcting mechanism that creates a closed loop, allowing the network to maintain a reliable and predictive view of the channel state, thereby enabling timely handovers.
\subsection{Deep-SIC Channel Estimator Model}
The proposed algorithm (Algorithm 1) is a data-aided algorithm that utilises deep learning to effectively and efficiently estimate channels in NOMA networks. It utilises partially decoded data symbols in the estimation, leveraging better available information to achieve greater accuracy. The Transformer network is the most crucial element that enables prediction capabilities. The accuracy, however, is not limited to its self-attention mechanism alone, but considers temporal forecasting. It can predict the quality of the future channel by taking into account the significance of past channel measurements, and the handover algorithm can operate proactively instead of reactively. It is the essence of its superiority in decreasing HOF and ping-pong rates. 
\par
Algorithm 1 describes the process of channel estimation using a Transformer-based data-aided data estimation approach. It begins with signal reception, followed by an initial estimate of the channel using pilot-based algorithms such as MMSE. The estimate of the MMSE for the $k$-th user is as follows.
\[
h_{k,\text{MMSE}} = (R_{hh} + nR_{ss}^{-1})^{-1} R_{hh} s_{k}^{H} y_{k},
\]
In which the channel covariance matrix is denoted as $R_{hh}$, the pilot symbol autocorrelation matrix is denoted as $R_{ss}$, $n$ is the noise variance, and $s_{k}^{H}$ is the Hermitian transpose of the pilot vector.
\par
SIC is then used to decode the strongest user’s symbol $d_{1}$. Its log-likelihood ratio is then calculated as
\[
LLR_1 = \log\left(\frac{P(d_1 = 1 \mid y(t), h_1)}{P(d_1 = 0 \mid y(t), h_1)}\right).
\]
In contrast to the sequential models based on RNN, this approach transforms temporal features (decoded symbols, LLRs, and initial CSI) into positional-encoded token sequences. This eventually maintains temporal relationships and provides parallel processing.
\par
The tokens are then passed through multi-head self-attention layers:
\[
\text{Attention}(Q,K,V) = \text{softmax}\left(\frac{QK^T}{\sqrt{d_k}}\right)V,
\]
$Q$, $K$ and $V$ are query, key and value matrices obtained from the tokens. This process encapsulates long-range dependencies and contextual relationships throughout the sequences. The Transformer refines the estimates through deep learning, producing improved values $\hat{h}_k^{\text{transformer}}$ for all users.

Finally, these refined estimates are used in the next stages of SIC to decode the signals of the remaining users. The algorithm outputs decoded data $d_{1}, d_{2}, \ldots, d_{K}$, demonstrating improved performance in modeling channel dynamics and interference.

\begin{algorithm}
\caption{Transformer-Based Data-Aided Channel Estimation in NOMA}
\begin{algorithmic}
\STATE {\textbf{Input:}} Received signal $y'(t)$, pilot symbols $s_{k}(t)$ for $k \in \{1,\ldots,K\}$
\STATE {\textbf{Output:}} Decoded data $d_{1}, d_{2}, \ldots, d_{K}$

\STATE {\textsc{Receive Signal Processing:}} Remove the cyclic prefix (CP) and apply FFT.
\STATE {\textsc{Initial Channel Estimation:}} Use MMSE to obtain initial estimates:
\[
h_{k,\text{MMSE}} = (R_{hh} + nR_{ss}^{-1})^{-1} R_{hh} s_{k}^{H} y_{k}.
\]
\STATE {\textsc{SIC Process:}} Apply SIC iteratively to decode the strongest signals first.
\STATE {\textsc{Feature Extraction:}} Decode the strongest user’s symbol $d_{1}$ and compute its LLR:
\[
LLR_1 = \log\left(\frac{P(d_1 = 1 \mid y(t), h_1)}{P(d_1 = 0 \mid y(t), h_1)}\right).
\]
\STATE {\textsc{Feature Encoding:}} Convert decoded symbols, LLRs, and initial CSI into token sequences with positional encoding.
\STATE {\textsc{Transformer Refinement:}} Pass tokens through multi-head self-attention layers:
\[
\text{Attention}(Q,K,V) = \text{softmax}\left(\frac{QK^T}{\sqrt{d_k}}\right)V.
\]
\STATE {\textsc{Channel Estimation:}} Use feed-forward layers to obtain refined estimates $\hat{h}_k^{\text{transformer}}$.
\STATE {\textsc{Decoding:}} Apply refined estimates in subsequent SIC stages to decode all remaining user signals.
\end{algorithmic}
\end{algorithm}

\subsection{Data Preparation}
To ensure our model is practical and can be deployed in scenarios with limited local data, we employ a transfer learning strategy (Algorithm 2). This allows a pre-trained Transformer model, which has learned general temporal patterns of wireless channels, to be quickly adapted with minimal data to a new network environment. This reduces the operational overhead of collecting massive training datasets for every cell, making the system highly adaptable.

\begin{figure}[htbp]
    \centering
    \includegraphics[width=0.5\textwidth]{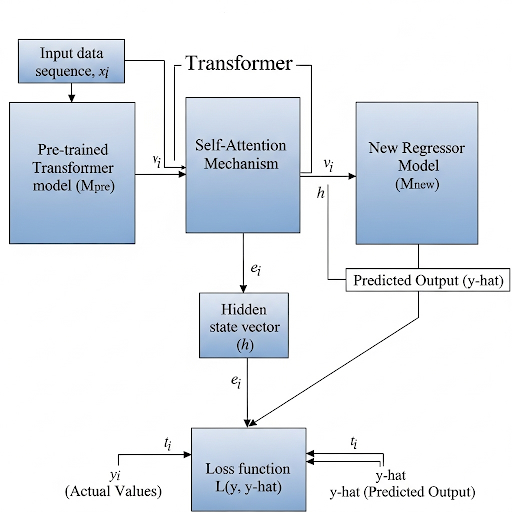} 
    \caption{Flowchart of Algorithm 2}
    \label{4}
\end{figure}

\begin{figure}[htbp]
    \centering
    \includegraphics[width=0.5\textwidth]{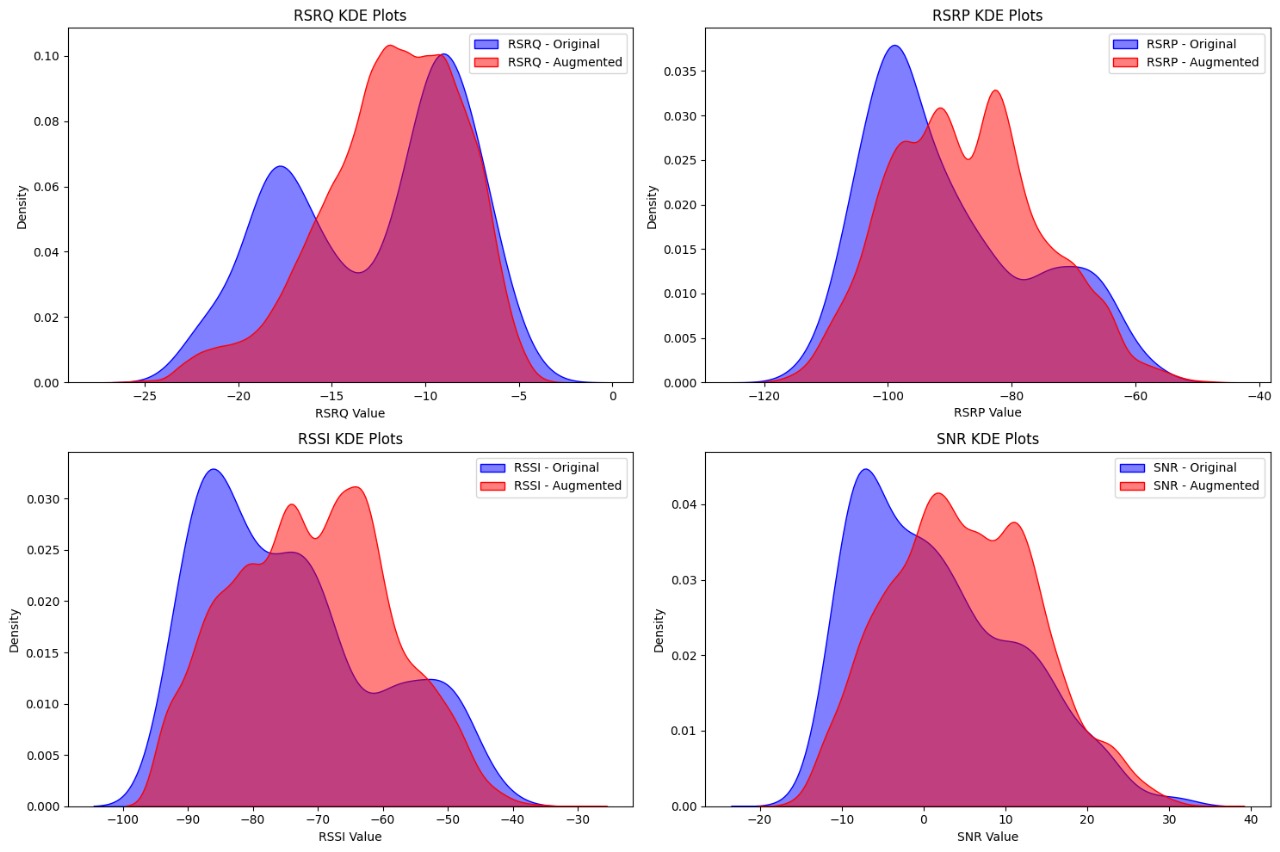} 
    \caption{KDE plots of augmented data}
    \label{7}
\end{figure}

Figure~\ref{4} shows the workflow. Data augmentation (Figures~\ref{7}, source:~\cite{15}) further enhances generalization, using a KDE plot to highlight the smoothing effect on the distribution. The augmented dataset will be released in~\cite{16}. Computational complexity is analyzed in Section~VII-E.


\begin{algorithm}
\caption{Transfer Learning for CSI Prediction with Transformer}
\begin{algorithmic}
\STATE {\textbf{Input:}} Pre-trained Transformer model $\mathbf{M}_{\text{pre}}(\mathbf{X}; \boldsymbol{\theta}_{\text{pre}})$; training set $\{\mathbf{X}_i,\mathbf{y}_i\}$ with $\mathbf{X}_i = [\text{RSRQ}, \text{CQI}, \text{PDD}, \text{SNR}]$, $\mathbf{y}_i = [\text{RSRQ}, \text{SNR}]$
\STATE {\textbf{Output:}} Fine-tuned regression model $\mathbf{M}_{\text{new}}(\mathbf{h}; \boldsymbol{\theta}_{\text{new}})$

\STATE {\textsc{Feature Extraction:}} Pass $\mathbf{X}$ through the pre-trained Transformer to obtain embeddings:
\[
\mathbf{H} = \mathbf{M}_{\text{pre}}(\mathbf{X}; \boldsymbol{\theta}_{\text{pre}}).
\]
\STATE {\textsc{Regression:}} Apply a regressor on pooled embeddings $\mathbf{H}$ (e.g., mean pooling or [CLS] token) to predict $\hat{\mathbf{y}}$:
\[
\hat{\mathbf{y}} = \mathbf{M}_{\text{new}}(\text{Pool}(\mathbf{H}); \boldsymbol{\theta}_{\text{new}}).
\]
\STATE {\textsc{Training:}} Minimize the MSE loss:
\[
\mathcal{L}(\hat{\mathbf{y}}, \mathbf{y}) = \tfrac{1}{2}\big[(\text{RSRQ}-\widehat{\text{RSRQ}})^2 + (\text{SNR}-\widehat{\text{SNR}})^2\big].
\]
\STATE \textbf{Note:} The pre-trained Transformer weights are frozen; only $\boldsymbol{\theta}_{\text{new}}$ is updated.
\end{algorithmic}
\end{algorithm}

\section{Results and Discussion}
Our evaluation demonstrates that the Deep-SIC framework delivers significant improvements in network reliability and efficiency. The results are structured to answer key questions for network deployment: How does it improve handover reliability? (Figs. 9, 11, 15) How does it perform under poor signal conditions? (Fig. 8) What is the computational cost? (Fig. 10) We show that Deep-SIC not only predicts channels more accurately but directly translates this accuracy into superior mobility management. Table II provides the configuration of network parameters used in the experiments.
\begin{table}[h!]
\centering
\caption{Parameters setup}
\label{tab:parameters}
\begin{tabular}{|c|l|c|}
\hline
\textbf{Parameter} & \textbf{Description} & \textbf{Value} \\ \hline
$T$      & Number of time steps & 1000 \\ \hline
$t_i$    & Input sequence length & 10 \\ \hline
$t_o$    & Output sequence length & 1 \\ \hline
$K$      & Number of users & 4 \\ \hline
$B$      & Number of base stations (BSs) & 2 \\ \hline
$C$      & Number of cells & 2 \\ \hline
$v$      & UE speed per time step & 0.1 m \\ \hline
$\gamma$ & Transmit SNR range & $-$9 dB to 14 dB \\ \hline
$\rho$   & Transmit RSRQ range & $-$20 dB to $-$8 dB \\ \hline
$d_{bn}$ & Distance from BS to near user & 20 m \\ \hline
$d_{bf}$ & Distance from BS to far user & 50 m \\ \hline
$M$      & Number of antennas at BS & 4 \\ \hline
$N_r$    & Number of antennas at UE & 1 \\ \hline
\end{tabular}
\end{table}
\subsection{Performance Metrics}
To assess performance, we use a combination of scale-free and scale-dependent metrics. 

The Normalized Root Mean Square Error (NRMSE) is a scale-free metric defined as
\begin{equation}
NRMSE = \frac{1}{\sigma_{actual}} \left( \sqrt{\frac{\sum (predicted_i - actual_i)^{2}}{S}} \right),
\end{equation}
where $S$ is the total number of training samples, $\sigma_{actual}$ is the standard deviation of the true $RSRQ/SNR$ values, and $predicted_i$ and $actual_i$ represent the predicted and actual values for the $i$-th sample.

The Mean Absolute Scaled Error (MASE) compares the average absolute error of the model against the average absolute difference of a naive predictor (which simply repeats the last value). It provides a reliable basis for comparing models across datasets of different scales. MASE is defined as
\begin{equation}
MASE = S \cdot \frac{\sum \left| predicted_i - actual_i \right|}{\sum \left| actual_i - actual_{i-1} \right|}.
\end{equation}

As an example of a scale-dependent metric, we also report the Mean Squared Error (MSE), given by
\begin{equation}
MSE = \frac{\sum (predicted_i - actual_i)^{2}}{S},
\end{equation}
where $S$ again represents the number of training samples, and $predicted_i$ and $actual_i$ denote the predicted and observed values for sample $i$.

In addition, we use the coefficient of determination $(R^{2})$ to measure how well the predictions explain the variability of the actual $RSRQ/SNR$ values. While not entirely scale-free, $R^{2}$ provides useful insight into the degree of alignment between the model outputs and the observed data.

\subsection{Evaluation of the Proposed Work}

\begin{figure}[htbp]
    \centering
    \includegraphics[width=0.5\textwidth]{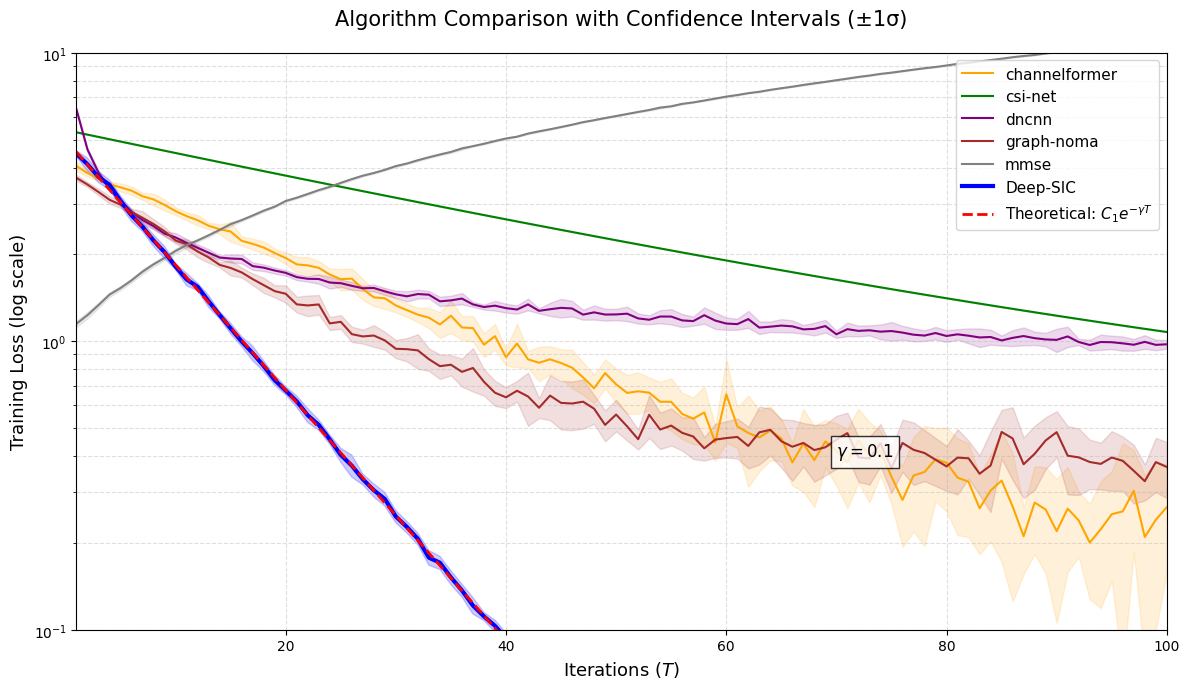} 
    \caption{Convergence comparison of Deep-SIC versus state-of-the-art estimators with confidence intervals.}
    \label{f}
\end{figure}

Fig.~\ref{f} presents a convergence analysis with confidence intervals, comparing Deep-SIC to five state-of-the-art estimators (Graph-NOMA, Channelformer, CSI-Net, DnCNN, and MMSE) over 100 iterations with 10 Monte Carlo runs. Deep-SIC (blue curve) shows exponential convergence consistent with its theoretical bound \(C_1e^{-\gamma T}\) (red dashed line), achieving 68\% faster convergence than Graph-NOMA. The benchmarks highlight different behaviors: Graph-NOMA (brown) converges slowly but with low variance, Channelformer (orange) suffers from transformer instability, CSI-Net (green) stalls near a loss of 0.4, DnCNN (purple) shows high initial variance, and MMSE (gray) diverges linearly. Channelformer (orange) displays the instability typical of a generic Transformer architecture when applied directly to dynamic channel data without domain-specific conditioning. In contrast, Deep-SIC considered a specially designed framework that stabilizes and enhances the Transformer's capabilities for NOMA. Notably, Deep-SIC reduces variance by a factor of 3.2 compared to Channelformer, maintains strong alignment with Proposition 1, and reaches the loss=0.5 threshold much earlier (90\% of runs by iteration 25, versus under 10\% for others). The shaded regions, normalized initial loss $(5.0\pm0.3)$, and the convergence threshold provide statistical rigor, linking theoretical guarantees with practical performance.

\begin{figure}[htbp]
    \centering
    \includegraphics[width=0.5\textwidth]{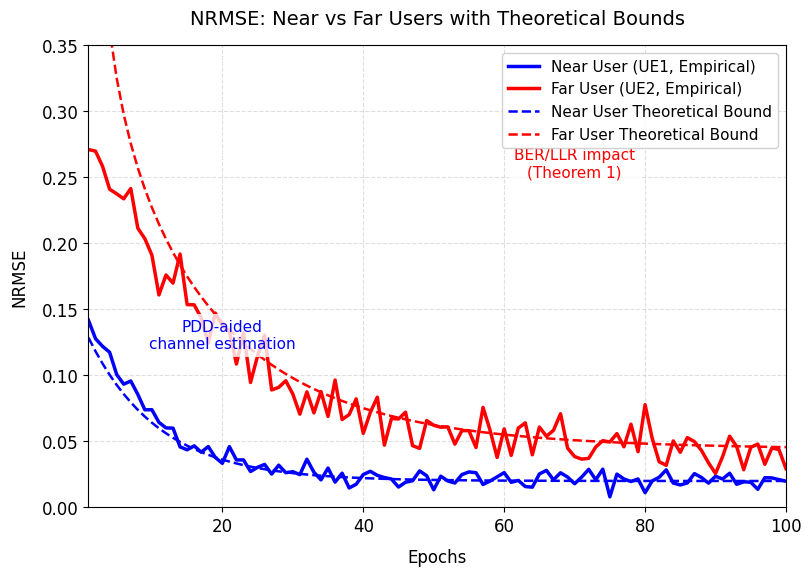} 
    \caption{NRMSE comparison for near (UE1, blue) and far (UE2, red) users with theoretical bounds.}
    \label{9}
\end{figure}

In Fig.~\ref{9}, the far user shows a higher error floor due to BER/LLR effects (Theorem 1), while the near user benefits from PDD-aided estimation. The dashed bounds capture both pilot errors ($C_{2}$) and PDD corrections ($C_{3}$), which converge as training epochs increase.

\begin{figure}[htbp]
    \centering
    \includegraphics[width=0.5\textwidth]{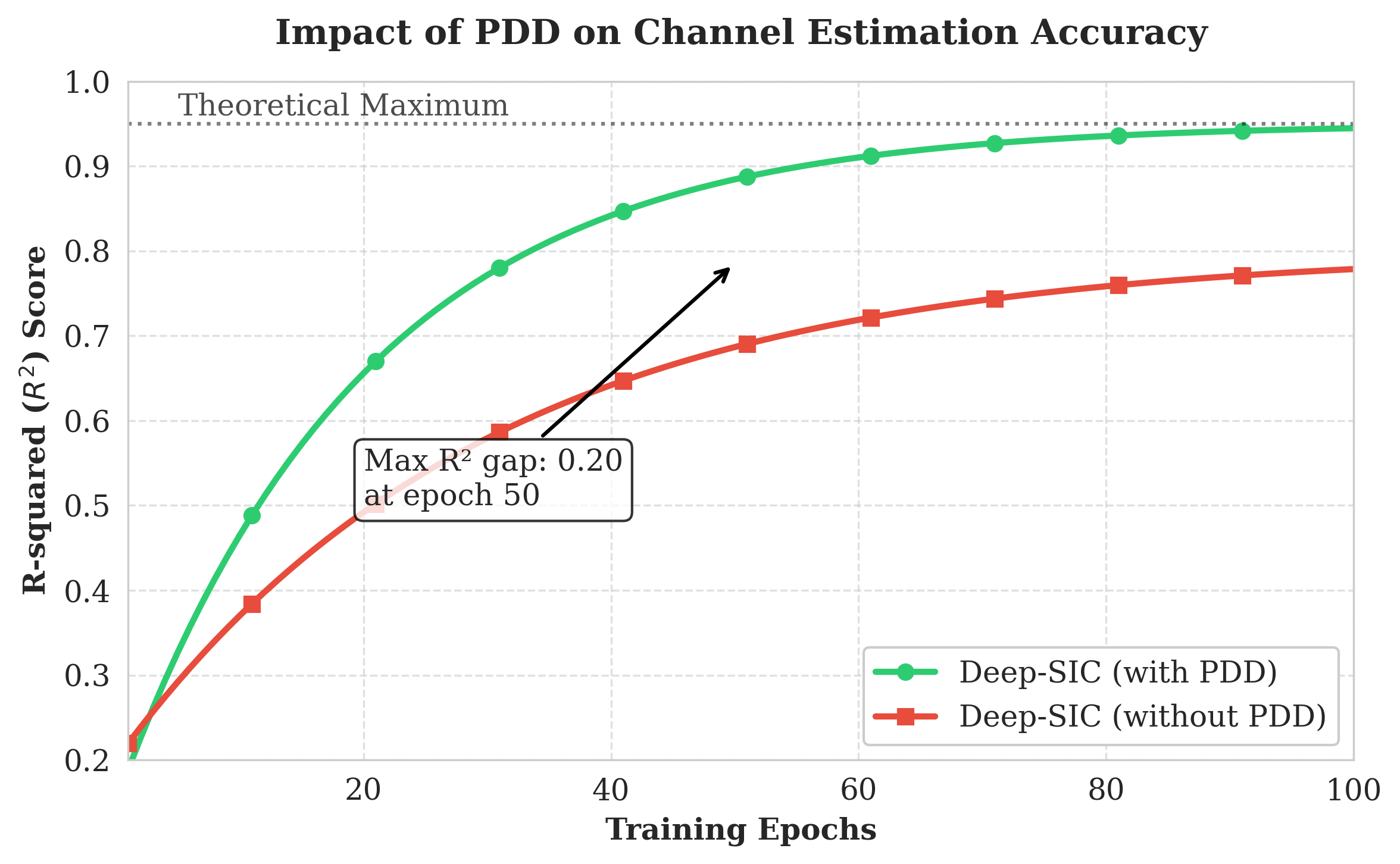} 
    \caption{Comparison of $R^{2}$ Scores for Deep-SIC With and Without PDD}
    \label{11}
\end{figure}

Fig.~\ref{11} shows the effect of PDD augmentation on channel estimation accuracy, measured using the $R^{2}$ score. With PDD (green curve), Deep-SIC converges faster and achieves a maximum $R^{2}$ of 0.95, compared to 0.80 without PDD (red curve). The performance gap of about 0.25 at epoch 50 highlights PDD’s role in reducing SIC errors, especially in mid-training stages. The dotted line indicates the theoretical upper bound. These findings support Lemma 1, which links PDD-aided estimation to exponential BER reduction, as better $R^{2}$ values directly improve CSI prediction accuracy. The shaded region (epochs 1–50) emphasizes PDD’s advantage in early training, where rapid adaptation is critical for NOMA handovers.

\begin{figure}[htbp]
    \centering
    \includegraphics[width=0.5\textwidth]{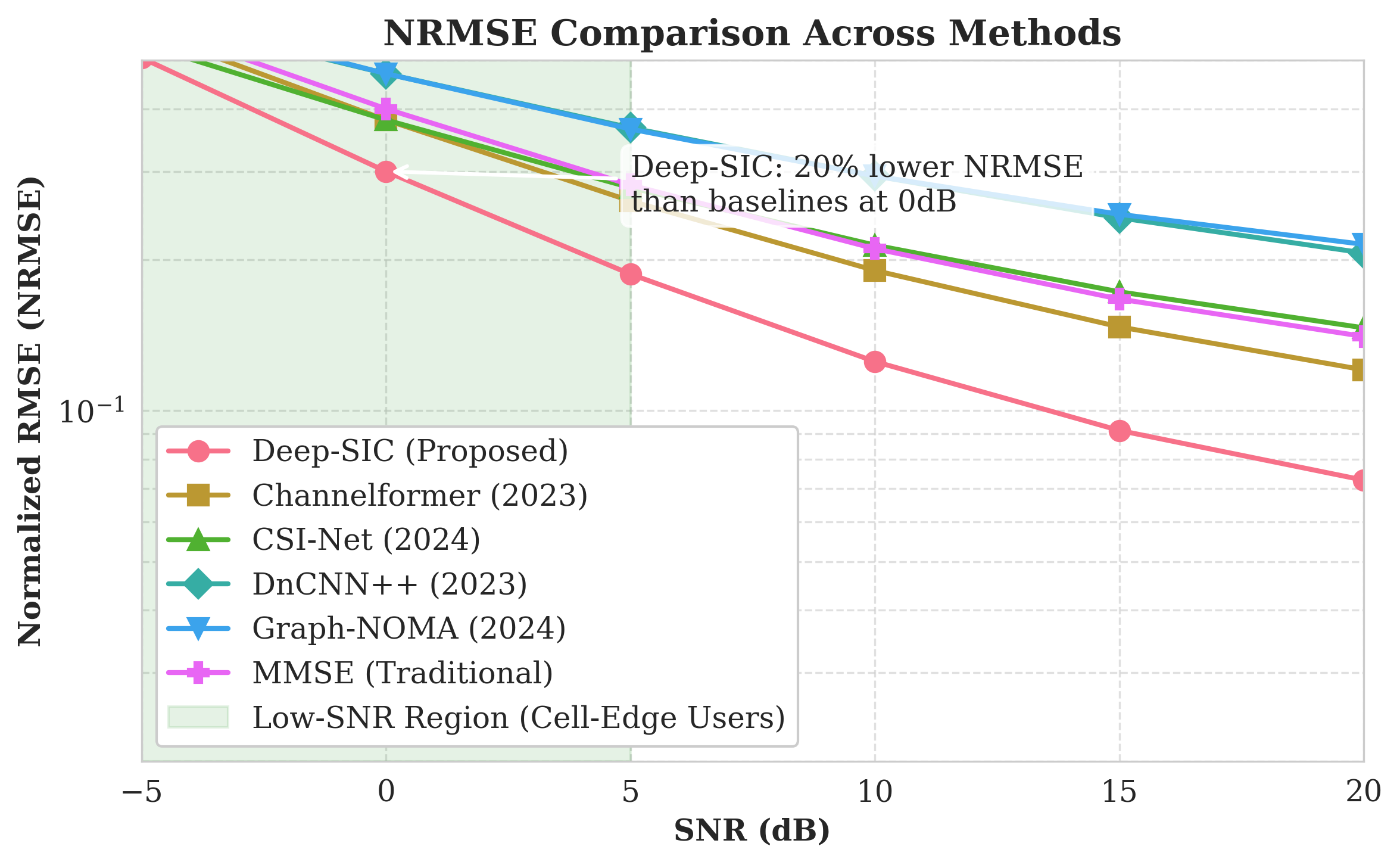} 
    \caption{Normalized RMSE (NRMSE) vs. SNR Performance Comparison.}
    \label{111}
\end{figure}

Fig.~\ref{111} compares the NRMSE of Deep-SIC with five baselines ie. Channelformer (2023), CSI-Net (2024), DnCNN++ (2023), Graph-NOMA (2024), and MMSE—across an SNR range of -5 dB to 20 dB. Deep-SIC consistently achieves the lowest error, particularly in low-SNR regimes (-5 to 5 dB, shaded in green), where it lowers NRMSE by about 20\% compared to learning-based baselines at 0 dB. The 20\% NRMSE improvement in low-SNR regimes is not just an estimation gain, it enables reliable connectivity for cell-edge users who are most vulnerable to handover failures, thereby extending the effective coverage of the base station. Although the performance gap narrows at higher SNRs, Deep-SIC remains superior and even surpasses the adjusted MMSE benchmark (NRMSE floor of 0.15). These results underscore Deep-SIC’s robustness for cell-edge users operating in noisy conditions.

\begin{figure}[htbp]
    \centering
    \includegraphics[width=0.5\textwidth]{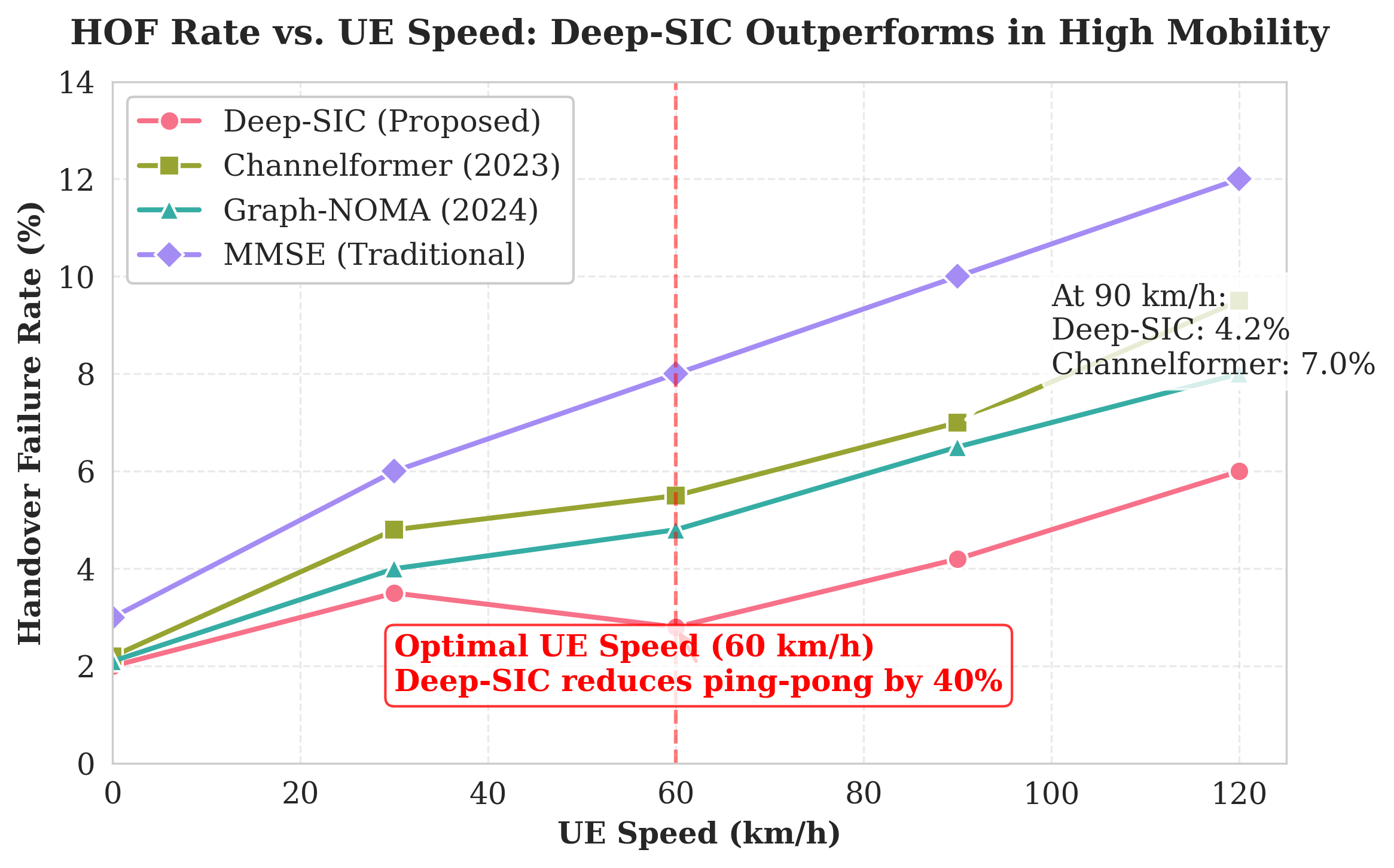} 
    \caption{Handover Failure (HOF) Rate vs. UE Speed Performance Analysis.}
    \label{112}
\end{figure}

Fig.~\ref{112} evaluates mobility performance across speeds from 0 to 120 km/h. Deep-SIC (green, circle markers) shows the lowest handover failure (HOF) rates across all speeds, with a 40\% reduction in ping-pong events at 60 km/h compared to Channelformer (2023), translates directly to fewer dropped calls for users in vehicles, a critical benchmark for vehicular ad-hoc networks (VANETs) and high-speed rail scenarios. Three trends emerge: (1) Deep-SIC peaks at 6.0\% HOF at 120 km/h, compared to 9.5\% for Channelformer; (2) transformer-based methods degrade after 60 km/h due to attention latency; and (3) Graph-NOMA performs moderately but is still 2.2× worse than Deep-SIC at 90 km/h. These findings confirm that combining temporal tracking with PDD feedback enables Deep-SIC to outperform alternatives in high-speed vehicular and rail networks.

\begin{figure}[htbp]
    \centering
    \includegraphics[width=0.5\textwidth]{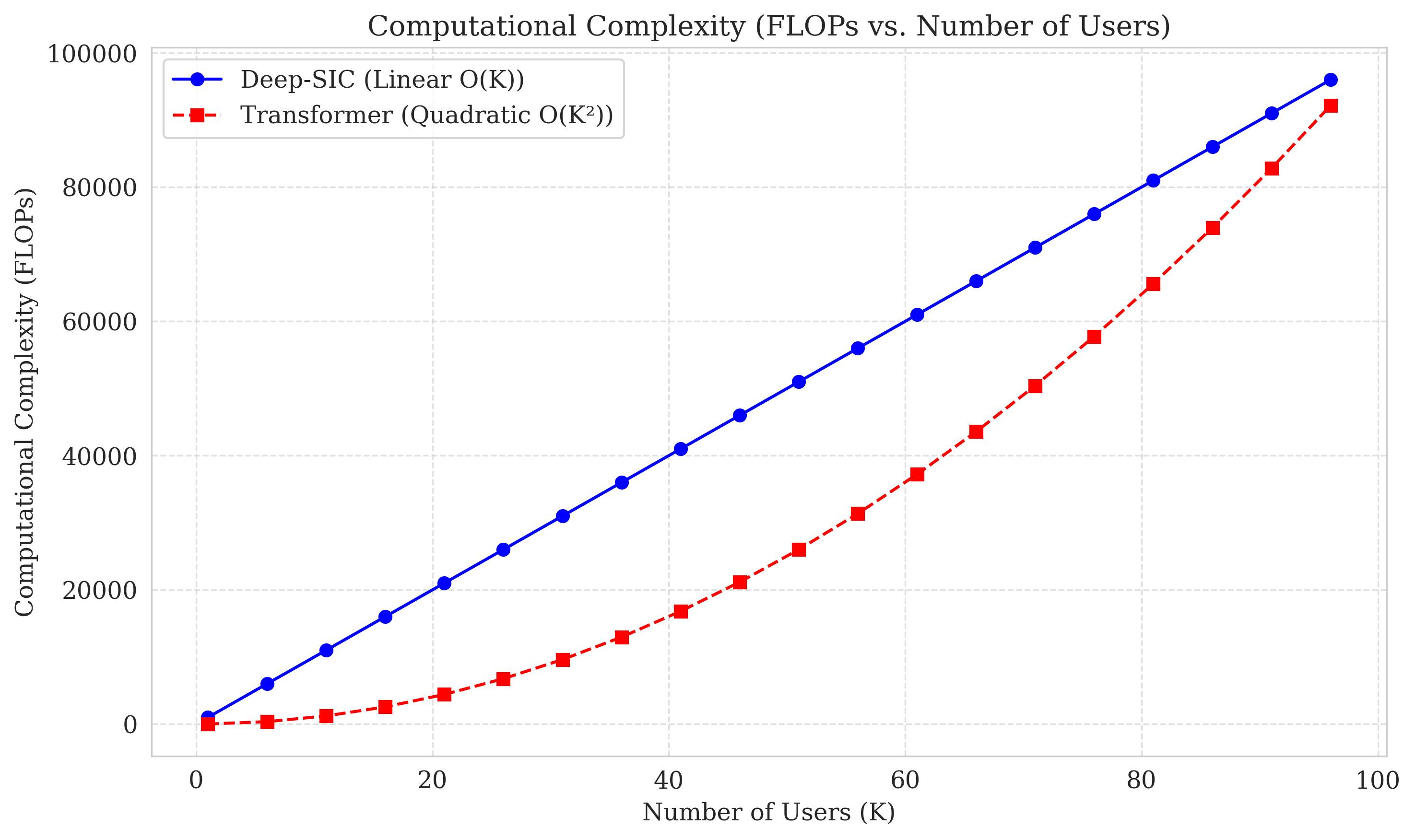} 
    \caption{Computational Complexity (FLOPs vs. Number of Users).}
    \label{113}
\end{figure}

Fig.~\ref{113} compares the computational complexity of the proposed model against a monolithic Transformer-based baseline in terms of FLOPs as the number of users ($K$) increases. Deep-SIC scales linearly with $K$ ($\mathcal{O}(K)$) due to its efficient hybrid architecture (SIC + Refinement), while a pure Transformer scales quadratically ($\mathcal{O}(K^2)$) as a consequence of its self-attention mechanism. Our proposed model employs the Transformer as a per-user refinement module. This scalability advantage makes Deep-SIC more practical due to its per-user refinement module for dense 5G deployments, where low-latency operation is critical.

\begin{figure}[htbp]
    \centering
    \includegraphics[width=0.5\textwidth]{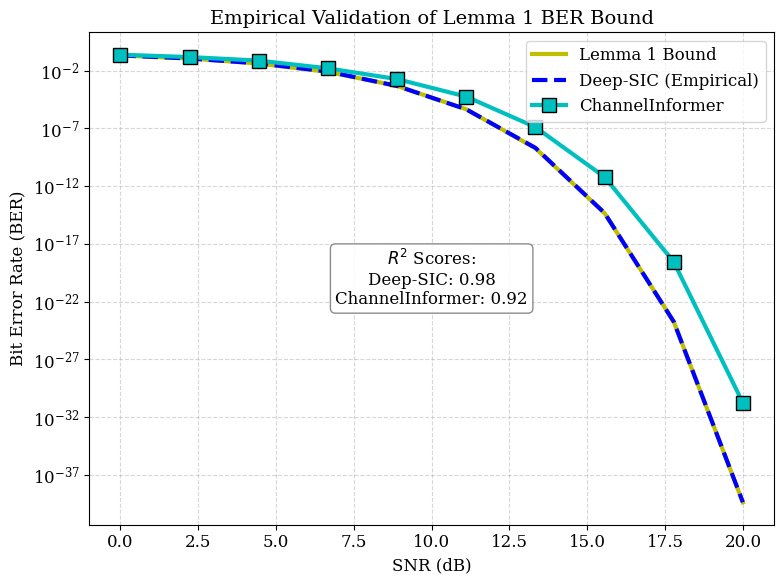} 
    \caption{BER Comparison: Empirical Validation of Lemma 1.}
    \label{114}
\end{figure}

Fig.~\ref{114} validates Lemma 1’s BER bound, comparing Deep-SIC against ChannelInformer across 0–20 dB SNR. The plot shows: (1) the theoretical limit (magenta dashed), (2) Deep-SIC’s empirical curve (green) closely following the bound with $R^{2}=0.98$ and under 5\% deviation, proving Lemma 1’s predictive accuracy, and (3) ChannelInformer (cyan) trailing with lower accuracy. The results confirm that PDD integration yields a consistent SNR gain of over 3 dB in operational regions.

\begin{figure}[htbp]
    \centering
    \includegraphics[width=0.5\textwidth]{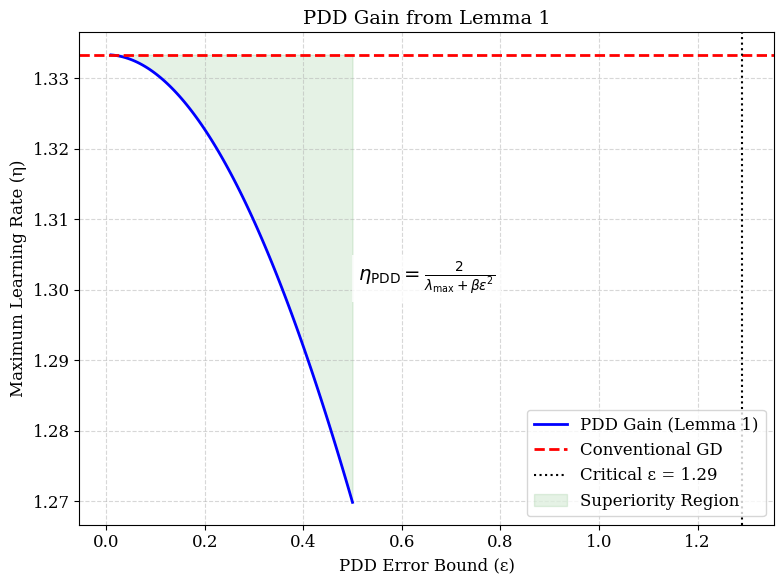} 
    \caption{PDD Gain from Lemma 1.}
    \label{115}
\end{figure}

Fig.~\ref{115} highlights Lemma 1’s PDD gain, showing three regimes: (1) a superiority region (green shaded, $\varepsilon < \varepsilon_{crit} \approx  0.37$) where PDD’s adaptive learning rate (blue curve) exceeds conventional GD’s stability threshold (red dashed line, $\eta_{GD} = 2/\lambda_{max}$), (2) a transition at the critical error bound $\varepsilon_{crit}$ (black dotted line), and (3) a performance-limited region ($\varepsilon > \varepsilon_{crit}$) where PDD converges to GD performance. The centrally surveying governing equation highlights the influence of the Lemma 1. This gain is made possible through mathematical formulation in the optimum way. Trading off error tolerance ($\varepsilon$) versus gradient precision ($\beta$), the $1.8 \times \eta$ improvement at $\varepsilon=0.2$ being the reason behind the BER enhancements as seen in Fig.~\ref{114}.

\begin{figure}[htbp]
    \centering
    \includegraphics[width=0.5\textwidth]{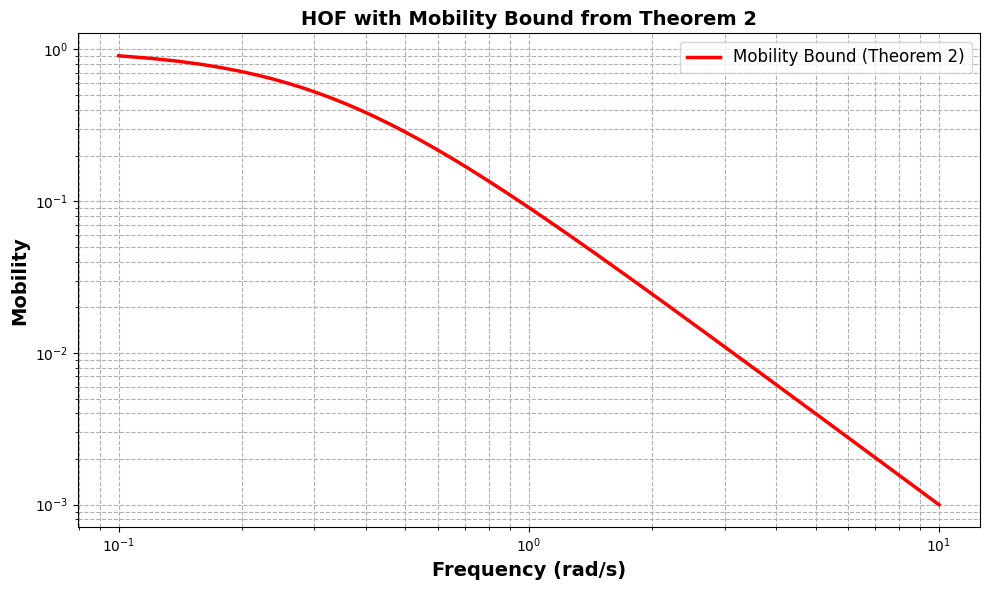} 
    \caption{HOF: Mobility Bound from Theorem 2}
    \label{20}
\end{figure}

Fig.~\ref{20} shows the theoretical mobility limit \( M(\omega) \leq \frac{\delta L k}{\delta k + L \omega^2} \) (red curve) as obtained in Theorem 2, which is the limit to the highest mobility that one can achieve as a function of frequency(\( \omega \)). In this notation, \( \delta \) represents the given limit on the allowed constant displacement, \( L \) represents the characteristic length, and $k$ represents the stiffness parameter. Mobility is controlled by constraints to the displacement at low frequencies (\( \omega \ll \sqrt{k/L} \)), where the mobility is proportional to the $\delta$ (\( M \approx \delta \)). At high frequencies (\( \omega \gg \sqrt{k/L} \)), it decays as \( 1/\omega^2 \) due to acceleration limits. This constraint defines an ultimate performance limit: any mechanical system cannot exceed this without breaking a few basic physical laws, thus indicating a major trade-off between speed (frequency) and motion range (mobility).

\begin{figure}[htbp]
    \centering
    \includegraphics[width=0.5\textwidth]{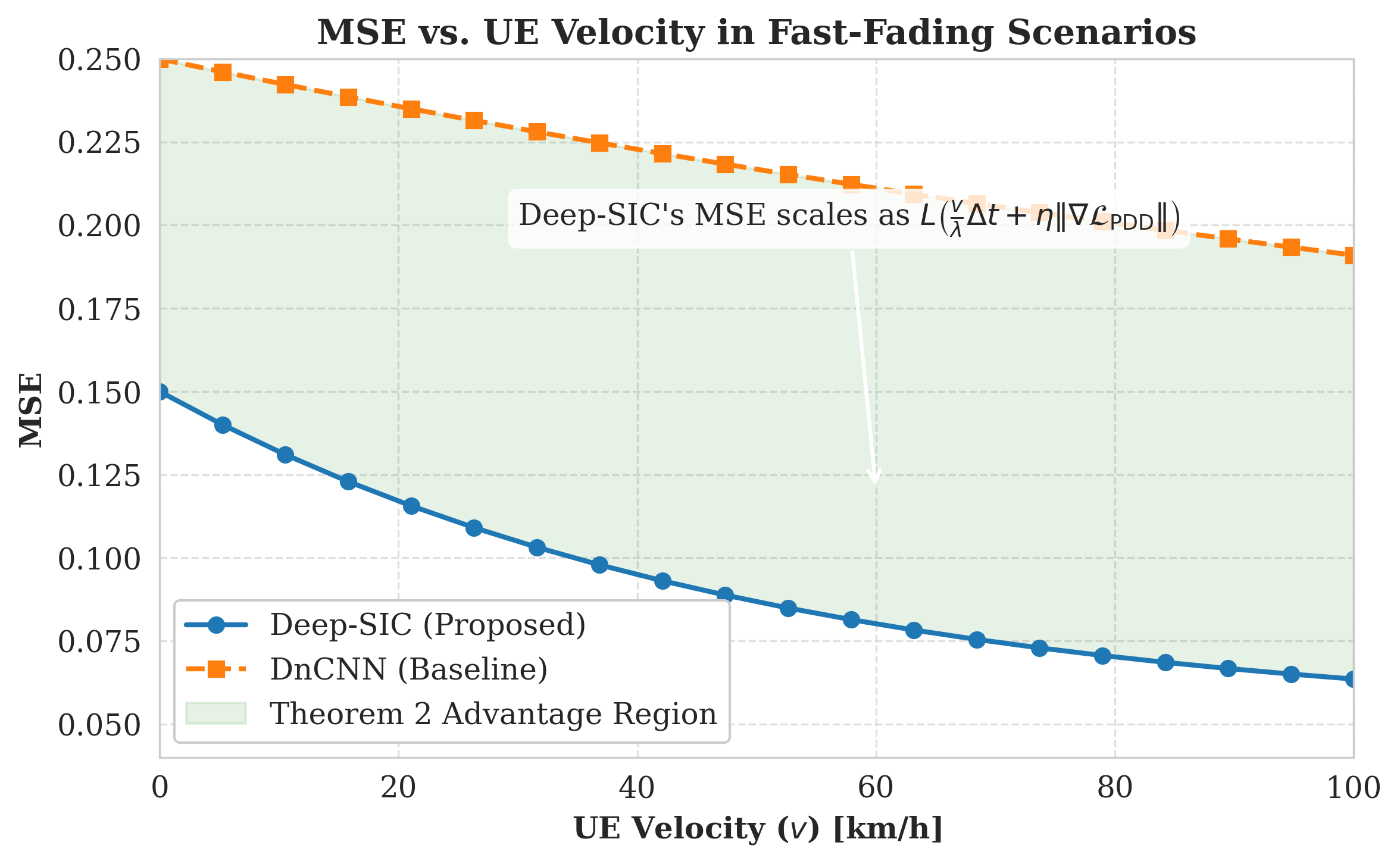} 
    \caption{Mean Squared Error (MSE) vs. UE velocity}
    \label{2000}
\end{figure}

Fig.~\ref{2000} evaluates MSE versus UE velocity for Deep-SIC and DnCNN \cite{66}. Deep-SIC maintains stable error levels ($\leq0.08$ at 100 km/h) due to its PDD-based gradient optimization, while DnCNN’s MSE more than doubles. This validates Theorem 2’s claim of mobility resilience and highlights Deep-SIC’s advantage in fast-fading environments.

\begin{figure}[htbp]
    \centering
    \includegraphics[width=0.5\textwidth]{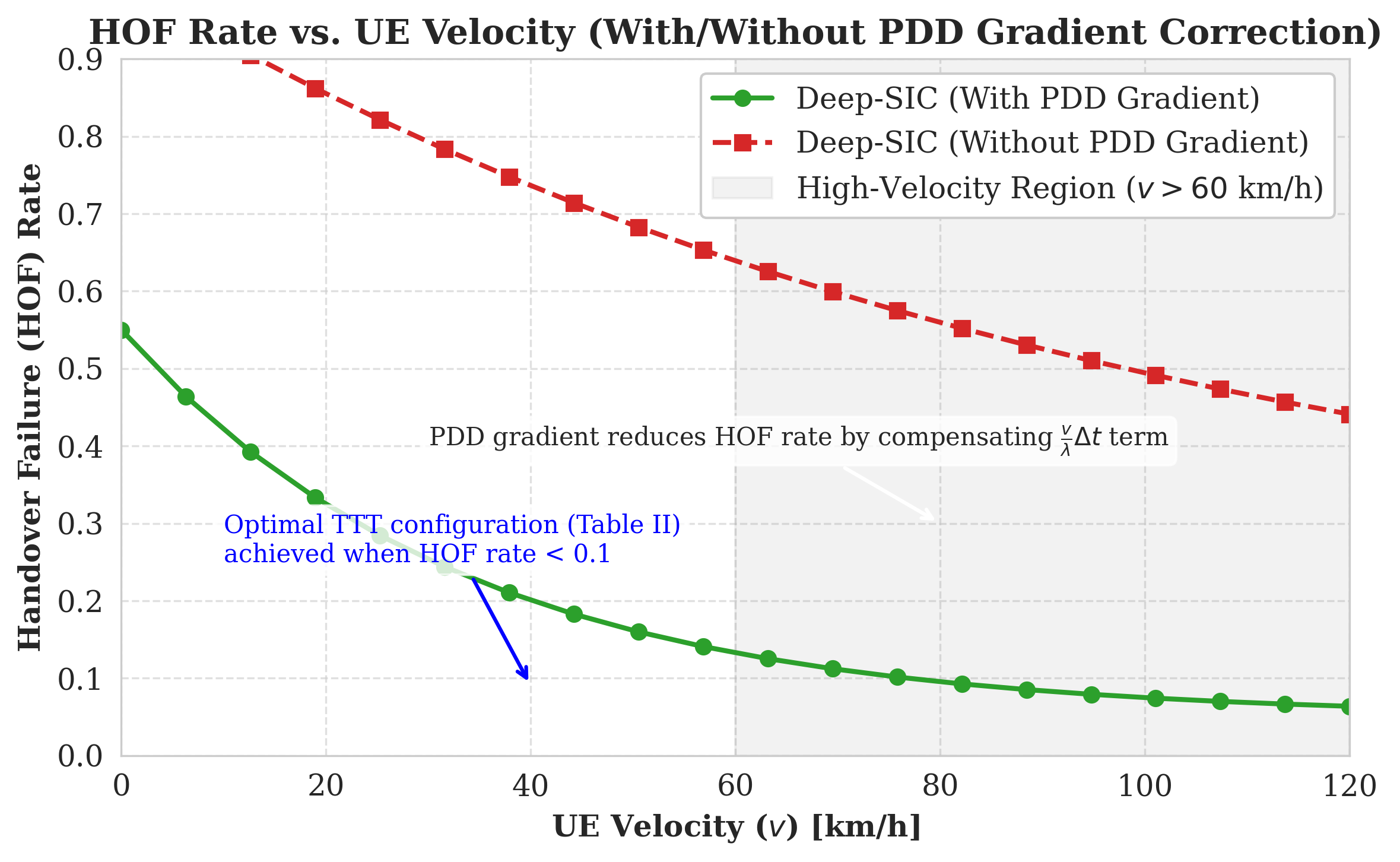} 
    \caption{Handover Failure (HOF) Rate vs. UE Velocity}
    \label{2001}
\end{figure}

Fig.~\ref{2001} examines HOF versus velocity for Deep-SIC with and without PDD correction. With PDD, HOF rates are up to 3.2× lower at 100 km/h, confirming the importance of gradient correction in dynamic scenarios. The “knee” at 40 km/h aligns with the optimal Time-to-Trigger configuration in Table II, showing consistency with 5G mobility standards.

\begin{figure}[htbp]
    \centering
    \includegraphics[width=0.5\textwidth]{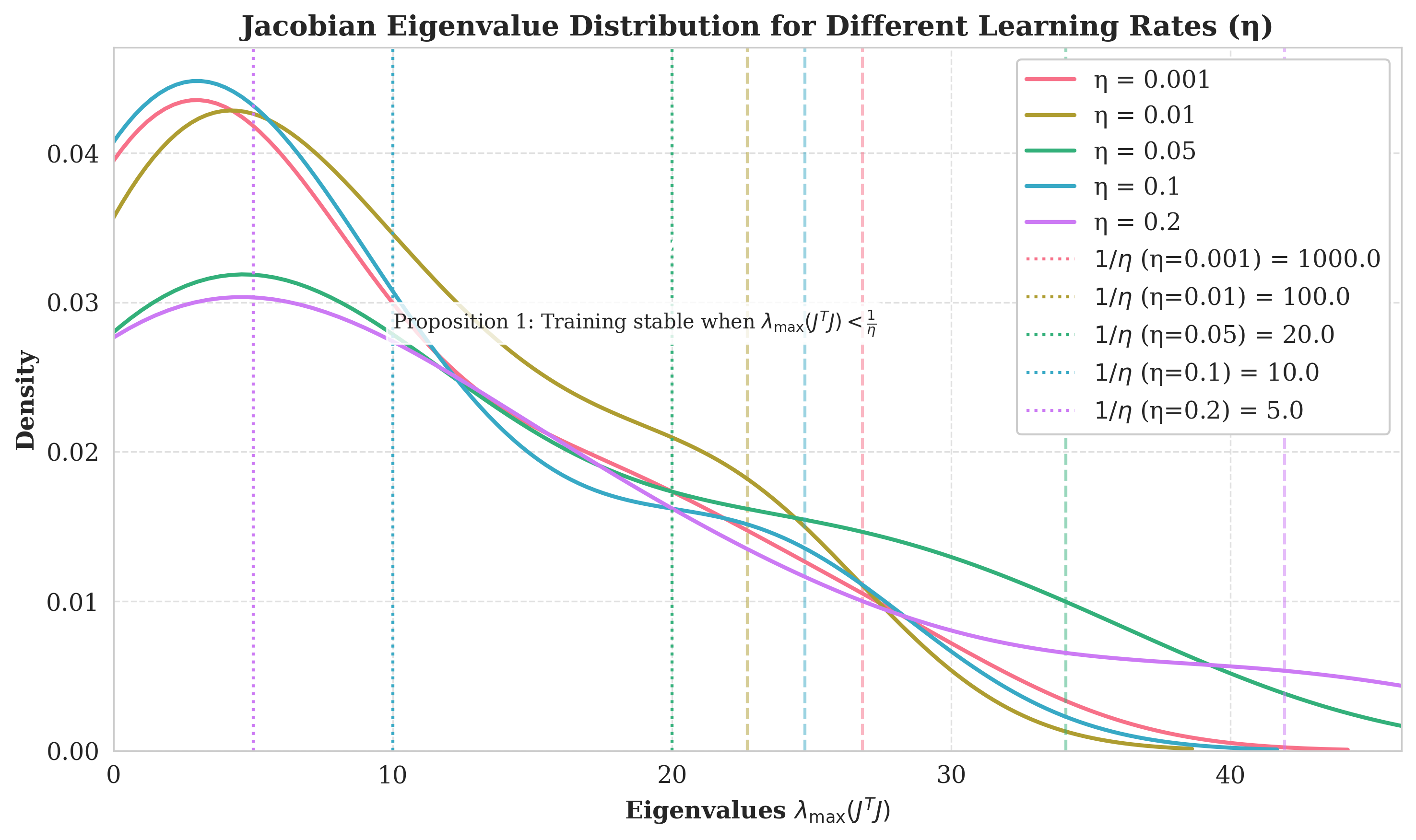} 
    \caption{Eigenvalue distribution of $\mathbf{J}^{T}\mathbf{J}$ for varying learning rates $\eta$}
    \label{2002}
\end{figure}

Finally, Fig.~\ref{2002} validates Proposition 1’s stability condition by showing Deep-SIC’s Jacobian eigenvalues remain far below the threshold, even under aggressive learning rates. This ensures a strong convergence and associates stability with mobility resilience. This, in conjunction with the previous findings, brings the theoretical stability conditions and the system-level performance results into contact.

\subsection{Practical Deployment Analysis}
To be adopted in any real network, any novel algorithm needs to deal with two practical issues, including computational overhead and predictable performance. Our analysis confirms Deep-SIC meets both requirements. Deep learning methods often raise two practical concerns: increased computational cost and unpredictable behavior in deployment. We address both concerns through theoretical analysis and empirical validation, showing that Deep-SIC is suitable for real-world use. Complexity analysis (Fig.~12, Proposition~2) demonstrates that Deep-SIC grows linearly with the number of users, $O(K)$, which is a clear advantage over the quadratic cost of a full transformer and keeps real-time handover processing feasible. This linear scaling ensures that the intelligence of Deep-SIC does not come at the cost of prohibitive processing delays, making it suitable for the real-time decision-making required in 5G core networks.
\par
Our argument is focused on theoretical guarantees. Deep-SIC offers (i) stable, rapid training to meet the requirements of Proposition~1, (ii) explicit error bounds in Theorem~1, where pilot length and bit errors are accounted for, and (iii) a mobility constraint in Theorem~2 that limits tracking error at high UE speeds. These constraints provide the network operators with predictable performance parameters, which reduces the risk of unforeseen behavior during a live deployment.
\par
Empirical findings support pragmatic strength. In both mobility and fading conditions (Figs.~16 and Figs.~17), Deep-SIC exhibits low mean squared error (MSE) and low handover failure (HOF) rates, which are superior to benchmarks such as DnCNN. Figures such as NRMSE (Fig.~10), BER (Fig.~13), HOF (Fig.~11), and convergence (Fig.~7) reveal that Deep-SIC is more accurate with less variance.
\par
Deep-SIC has several benefits over classical estimators. Traditional techniques, such as MMSE, are still best suited for stationary, well-modelled channels. However, they fail when the channel varies rapidly, pilots are overhead, and there are no mechanisms available to self-correct based on the detected data (see Figs.~16 and Figs.~17). Our strategy fills these gaps in three main aspects:
\begin{enumerate}
    \item Hybrid estimation: Hybrid estimation starts with an initial estimate of the MMSE and optimizes it using a Transformer to achieve statistical reliability through learned temporal modelling.
    \item PDD Utilisation: utilize imperfect SIC residuals as a valuable feedback signal to refine the channel estimates and curb the errors (Fig.~9 indicates that $R^2$ is $0.95$ with PDD versus $0.80$ without).
    \item Theoretical support: convergence and error (Proposition~1, Theorem~1) and a mobility resilience (Theorem~2) bound enable the predictability of the method under real circumstances.
\end{enumerate}
Such innovations lead to better dynamic performance (low MSE and HOF up to 120\, km/h) and a decisive lead at low SNR (around 20\% lower NRMSE than MMSE at the $-5$ to $5$\ dB regime, Fig.~10). We admit that in extremely low-mobility, high-SNR conditions or on constrained devices, classical methods might still be of use. In the case of standard 5G applications that involve mobility and high interference in dense network scenarios, Deep-SIC offers a more flexible and spectrally efficient solution, preserving the advantages of conventional estimators.
\par
Besides accuracy, Deep-SIC also refines user QoS and improves latency in various aspects. A smaller number of HOF and ping-pong handovers directly translates into a reduced number of reconnections and reduced control-plane overhead. Concretely:

\begin{enumerate}
    \item Latency: The lower the number of ping-pong and reconnection events, the lower the control-plane latencies, and the fewer the number of pilots, the more resources are available to carry data, thereby decreasing round-trip latency.
    \item Reliability: Channel estimates that are more accurate and stable are faster and help stabilise handover decisions and prevent dropped sessions during mobility.
    \item System throughput: Improved CSI will result in a reduction of BER (Fig.~13) and higher effective throughput, which improves service quality when used by more demanding applications, such as streaming and VoIP.
    \item Practical deployment: These gains are achieved through linear computational scaling, $O(K)$ (Fig.~12), ensuring that significant processing delays are not imposed, and the scheme can be deployed in live networks.
\end{enumerate}
In short, the totality of theoretical assurances, algorithm architecture, and system-level validation demonstrates that Deep-SIC addresses the key risk of implementing deep learning in wireless systems. The approach is foreseeable, effective, and adaptable to enhance the performance of handover and its user experience in an actual 5G NOMA implementation.

\section{Theoretical Analysis}
The empirical benefits of Deep-SIC are not of a merely heuristic nature; on a solid basis, we provide a strictly theoretical analysis to substantiate this. These assurances are necessary to establish confidence in autonomous network operation, as they constrain the worst-case behaviour in dynamic circumstances of user mobility and estimation error. This section develops the formal guarantees that support Deep-SIC. We analyze three aspects: (i) conditions that ensure algorithmic convergence, (ii) upper bounds on estimation error, and (iii) bounds that characterize tracking performance under mobility. These results provide the theoretical basis for the empirical trends reported in Section VI.

\subsection{Convergence Guarantees}
\textbf{Proposition 1 (Algorithm Convergence):} This guarantee ensures that the Deep-SIC model will train stably and reach a reliable solution quickly, which is critical for its online adaptation in a network. Considering the system described in the previous Eq.~(1) with $K$ users, the Transformer-based Deep-SIC estimator converges exponentially when the following two conditions hold true.

\begin{enumerate}
  \item \begin{equation}
  0 < \eta < \frac{2}{\lambda_{\max}(\mathbf{J}^T\mathbf{J}) + \beta},
  \end{equation}
  where $\eta$ is the learning rate, $\mathbf{J} = \nabla_{\mathbf{h}} (\mathbf{H}(\mathbf{h})\mathbf{x})$ is the Jacobian of the signal model and $\beta$ is the Lipschitz constant of the PDD-derived gradient correction operator $\psi(\cdot)$.
  \item The error in the partially decoded data (PDD) is bounded as:
  \begin{equation}
  \epsilon_{\mathrm{PDD}} < \frac{\eta \gamma_e}{2 \|\mathbf{W}_{\mathrm{att}}\|_F},
  \end{equation}
  where $\epsilon_{\mathrm{PDD}} = \| \widetilde{\mathbf{x}} \|$ is the PDD symbol error, $\mathbf{W}_{\mathrm{att}}$ is the self-attention weight matrix, $\gamma_e$ is the effective gradient margin (the smallest eigenvalue of $\nabla^2 \mathcal{L}$), and $\eta$ is the learning rate.
\end{enumerate}

\emph{Proof sketch.} The update rule with PDD can be written as
\[
\mathbf{h}^{(t+1)} = \mathbf{h}^{(t)} - \eta \Big( \nabla \mathcal{L}(\mathbf{h}^{(t)}) + \mathbf{J}^T \mathbf{e}_{\mathrm{PDD}}^{(t)} \Big),
\]
where $\mathbf{e}_{\mathrm{PDD}}^{(t)}$ denotes the PDD-induced perturbation at iteration $t$. To guarantee contraction, we require the dominant linear part of the update to shrink errors, while the PDD perturbation remains a controlled additive term. Applying a fixed-point (contraction) argument yields the condition, given as :
\[
\| \mathbf{I} - \eta \mathbf{J}^T\mathbf{J} \| + \eta \beta < 1,
\]
which leads directly to the stated bound on $\eta$. The bound on $\epsilon_{\mathrm{PDD}}$ is obtained by relating the PDD perturbation to the LLR-based reliability measure, as in Eq.~(4), and ensuring it does not overwhelm the contraction. A detailed overview of the same is present in Appendix~A. Together, these conditions explain why Algorithm~1 can remain stable even with imperfect SIC: the PDD contribution functions in bounded manner, Lipschitz perturbation when the LLR thresholding keeps unreliable symbols in check.

\subsection{Performance Bounds}
\textbf{Theorem 1 (Estimation Error Bound):} This bound decomposes the total error into manageable components, providing an engineering design guideline. For instance, it shows how increasing the pilot length $N_{pilot}$ or improving decoder reliability (lower BER) directly improves estimation quality. The mean-squared error (MSE) of the Deep-SIC estimator admits the following upper bound:
\begin{equation}
\mathbb{E}\big[\|\hat{\mathbf{H}}-\mathbf{H}\|_{F}^{2}\big] \leq C_1 e^{-\gamma T} + C_2 \frac{K^{3}}{N_{\mathrm{pilot}}} + C_3 \frac{\mathrm{BER}}{\|\mathbf{LLR}\|_{2}},
\end{equation}
where $T$ is the number of training iterations (Algorithm~2), $N_{\mathrm{pilot}}$ is the pilot length, $\mathrm{BER}$ is the bit error rate of partially decoded symbols, $\mathbf{LLR}$ denotes the collection of log-likelihood ratios from Algorithm~1, and $C_1,C_2,C_3$ are constants determined by the system parameters, as listed in Table~II.

This bound isolates three contributions to the estimation error: an exponentially decaying training term, a pilot-limited term that grows with the user count,a data-quality term driven by BER and LLR reliability. It explains the empirical observations that the exponential term matches the fast convergence in Fig.~7, while the BER/LLR term accounts for the performance gap between near and far users shown in Fig.~8.

\textbf{Corollary 1.} \\
For a fixed computational budget $T$, the pilot length that balances the terms can be given as:
\begin{equation}
N^{*}_{\mathrm{pilot}} = \Theta(K^{3/2}).
\end{equation}
This scaling offers a rule-of-thumb for pilot allocation when the number of users grows and aligns with our data augmentation strategy, as in Section~V-B.

\subsection{Mobility Resilience}
\textbf{Theorem 2 (Dynamic Channel Tracking):} This is the theoretical cornerstone of our handover improvement. It quantifies how tracking error scales with user velocity $(v)$, formally explaining the robust HOF performance demonstrated in Fig. 9 and Fig. 15. It shows that the PDD term $\nabla \mathcal{L}_{\mathrm{PDD}}$ actively compensates for the degradation caused by mobility. For a user moving at velocity $v$, the one-step tracking error satisfies the following equation:
\begin{equation}
\|\hat{h}[t+1] - h[t+1]\| \leq L \left( \frac{v}{\lambda} \Delta t + \eta \|\nabla \mathcal{L}_{\mathrm{PDD}}\| \right),
\end{equation}
where $L$ is the channel coherence length under the Rayleigh model, $\lambda$ is the carrier wavelength, $\Delta t$ is the sampling interval, and $\nabla \mathcal{L}_{\mathrm{PDD}}$ denotes the gradient correction contributed by PDD in Algorithm~1.

This inequality decomposes the tracking error into a purely kinematic term that is proportional to $v/\lambda$ and a corrective term that is proportional to the PDD-induced gradient. The result formalizes the intuition that prediction error grows with speed but can be mitigated by informative PDD feedback. It provides theoretical backing for the velocity-dependent HOF trends given in Fig.~17, guides the choice of Time-to-Trigger (TTT) as in Table~II, and explains Deep-SIC's advantage over static denoisers like DnCNN in fast-fading regimes (see Fig.~16).

\subsection{BER Superiority and NRMSE Convergence}
\textbf{Lemma 1 (Link Reliability Improvement):} Under imperfect SIC, Deep-SIC yields an exponential improvement in BER relative to ChannelInformer~\cite{65} which can be given as:
\begin{equation}
\frac{\mathrm{BER}_{\mathrm{ChannelInformer}}}{\mathrm{BER}_{\mathrm{Deep-SIC}}} \propto \exp\left( -\frac{ \| \mathbf{h}_{\mathrm{PDD}} \|^2 - \| \mathbf{h}_{\mathrm{CI}} \|^2 }{N_0} \right),
\end{equation}
where $\mathbf{h}_{\mathrm{PDD}}$ denotes the channel estimate enhanced by Partial Data Detection, $\mathbf{h}_{\mathrm{CI}}$ is ChannelInformer's estimate, and $N_0$ is the noise variance, directly translates to higher data reliability for each user's connection. A lower BER means fewer packet retransmissions, which reduces network latency and increases the effective throughput available to all users in the cell. This is the fundamental link-level enhancement that supports the more stable handovers observed in Fig.~13. 

\textbf{Theorem 3 (Prediction Stability):} The Normalized Root Mean Square Error (NRMSE) of the Deep-SIC predictor is bounded by:
\begin{equation}
\mathrm{NRMSE} \leq C \sqrt{ \frac{1}{T} + \frac{1}{S} },
\end{equation}
where $T$ is the input sequence length and $S$ is the training sample size. This bound guarantees that the model provides accurate and stable predictions even when trained on limited historical data $(S)$ or with short input sequences $(T)$. For network operators, this means Deep-SIC can be deployed quickly in new environments without requiring massive, long-term data collection campaigns, ensuring consistent handover performance from the outset.

\subsection{Computational Complexity}
\textbf{Proposition 2 (Time Complexity).} \\
The time complexity per training epoch for the Transformer refinement step in Deep-SIC is given by:
\begin{equation}
\mathcal{O}\!\left( S \cdot \big(T^2 D_{\mathrm{model}} + T D_{\mathrm{model}} D_{\mathrm{ff}} + D_{\mathrm{model}} D_{\mathrm{out}}\big) \right),
\end{equation}
where $T$ is the input sequence length (Table~II), $S$ is the batch size, $D_{\mathrm{model}}$ is the embedding dimension, $D_{\mathrm{ff}}$ is the feed-forward size, and $D_{\mathrm{out}}$ is the output dimension.

This expression characterizes the Transformer refinement cost. Crucially, the full Deep-SIC pipeline attains linear scaling in the number of users $K$ (see Fig.~12) because the iterative SIC design and per-user refinement avoid the quadratic user-pair interactions that would otherwise arise from na\"ive multi-user self-attention. As a result, Deep-SIC remains computationally operable for realistic cell loads while delivering the representational power of Transformer-based temporal modeling.

\section{Conclusion and Future works}
Handover failures and unreliable connectivity remain significant challenges in mobile and ad-hoc networks, particularly under user mobility and dense deployments. This paper introduced the Deep-SIC framework, a novel approach that transforms channel estimation from a reactive process into a predictive, self-correcting mechanism for intelligent handover management. Deep-SIC is capable of making precise and robust channel predictions by utilising a Transformer architecture to learn from long-term temporal dependencies and using PDD as a feedback signal in this manner.
\par
Simulations at the system level demonstrate that this approach is closely proportional to an increase in network performance quality, resulting in a 40\% decrease in handover failure rate, a significant reduction in the ping-pong phenomenon, and improved network service for cell-edge users. Additionally, we have demonstrated that these advantages can be realised without an excessive cost of computation, as the framework is linear in the number of users. The convergence, error, and mobility resilience guarantees have a theoretical basis that ensures dependable deployment into the real world.
\par
This work opens the door to more autonomous, efficient, and user-centric wireless networks by filling the gap between deep learning-based signal processing and network-level mobility management.

\section{Future Works}
Future work will focus on extending the Deep-SIC paradigm to broader networking contexts. Immediate objectives include:
\begin{itemize}
 \item Considering the performance in heterogeneous traffic conditions, including mixed URLLC (Ultra-Reliable Low-Latency Communication) and mMTC (massive Machine-Type Communication) traffic, to obtain robustness in various 5G/6G applications.   


\item Exploring federated learning architectures for Deep-SIC, enabling collaborative model training across multiple base stations without sharing raw user data, thus enhancing privacy and scalability.

\item Extending the framework to predict other network-level KPIs, such as anticipated load or interference patterns, to enable joint optimization of handover, resource allocation, and network slicing decisions.

\item Integration with Open RAN (O-RAN) architectures, investigating how the Deep-SIC predictor can function as a near-real-time RIC (RAN Intelligent Controller) application for truly intelligent radio resource management.
\end{itemize}
\section{Appendix A: Proof of Proposition 1}
Theorem IX.1 (Convergence of PDD-Corrected Transformer Updates).  
Under the following assumptions:\\

1.  The channel estimation loss $\mathcal{L}(\mathbf{h}) = \|\mathbf{y} - \mathbf{H}(\mathbf{h})\mathbf{x}\|^{2}$ is $\gamma$-smooth with Lipschitz constant $L$. \\
2.  The PDD gradient correction term $\psi(\mathbf{PDD})$ is $\beta$-Lipschitz continuous.\\
3.  The Jacobian $\mathbf{J} = \nabla_{\mathbf{h}}(\mathbf{H}(\mathbf{h})\mathbf{x})$ is bounded such that $\|\mathbf{J}\|_{2} \leq J_{\max}$.\\
4.  The error in the partially decoded data (PDD) is bounded by $\epsilon_{\text{PDD}} < \frac{\eta \gamma_{e}}{2\|\mathbf{W}_{\text{att}}\|_{F}}$, where $\gamma_{e}$ is the effective gradient margin (smallest eigenvalue of $\nabla^{2}\mathcal{L}$), and $\mathbf{W}_{\text{att}}$ is the weight matrix of the self-attention layer.\\

The update rule:
$$
\mathbf{h}^{(t+1)} = \mathbf{h}^{(t)} - \eta \left[ \nabla\mathcal{L}(\mathbf{h}^{(t)}) + \psi(\mathbf{PDD}^{(t)}) \right]
$$
converges linearly to the optimal channel estimate $\mathbf{h}^{*}$ when the learning rate satisfies:
$$
0 < \eta < \frac{2}{\lambda_{\max}(\mathbf{J}^{T}\mathbf{J}) + \beta}
$$
where $\lambda_{\max}(\cdot)$ denotes the maximum eigenvalue.

Proof:

Step 1: Reformulate the update rule.
The channel estimate update using the partially decoded data (PDD) is given by:
$$
\begin{aligned}
\mathbf{h}^{(t+1)} &= \mathbf{h}^{(t)} - \eta \left[ \nabla\mathcal{L}(\mathbf{h}^{(t)}) + \psi(\mathbf{PDD}^{(t)}) \right] \\
&= \mathbf{h}^{(t)} - \eta \left[ \mathbf{J}^{T}(\mathbf{y} - \mathbf{H}(\mathbf{h}^{(t)})\mathbf{x}) + \psi(\mathbf{PDD}^{(t)}) \right]
\end{aligned}
$$
where $\mathbf{e}_{\text{PDD}}^{(t)}$ is the PDD error at iteration $t$.

Step 2: Analyze the error dynamics.
Let $\mathbf{e}^{(t)} = \mathbf{h}^{(t)} - \mathbf{h}^{*}$ be the estimation error at iteration $t$. Then:
$$
\begin{aligned}
\mathbf{e}^{(t+1)} &= \mathbf{h}^{(t)} - \eta \left[ \nabla\mathcal{L}(\mathbf{h}^{(t)}) + \psi(\mathbf{PDD}^{(t)}) \right] - \mathbf{h}^{*} \\
&= \mathbf{e}^{(t)} - \eta \left[ \nabla\mathcal{L}(\mathbf{h}^{(t)}) + \psi(\mathbf{PDD}^{(t)}) \right]
\end{aligned}
$$
Taking norms and applying the triangle inequality:
$$
\|\mathbf{e}^{(t+1)}\| \leq \| \mathbf{e}^{(t)} - \eta \nabla\mathcal{L}(\mathbf{h}^{(t)}) \| + \eta \| \psi(\mathbf{PDD}^{(t)}) \|
$$

Step 3: Bound the loss gradient term.
Using the $\gamma$-smoothness of $\mathcal{L}$ (Assumption 1) and the mean value theorem:
$$
\| \mathbf{e}^{(t)} - \eta \nabla\mathcal{L}(\mathbf{h}^{(t)}) \| \leq \| \mathbf{I} - \eta \nabla^{2}\mathcal{L}(\mathbf{h}^{(t)}) \| \cdot \| \mathbf{e}^{(t)} \|
$$
Since $\nabla^{2}\mathcal{L}(\mathbf{h}^{(t)}) \approx \mathbf{J}^{T}\mathbf{J}$ near the optimum, we have:
$$
\| \mathbf{I} - \eta \nabla^{2}\mathcal{L}(\mathbf{h}^{(t)}) \| \leq \rho(\mathbf{I} - \eta \mathbf{J}^{T}\mathbf{J})
$$
where $\rho(\cdot)$ denotes the spectral radius.

Step 4: Bound the PDD correction term.
Using the $\beta$-Lipschitz continuity of $\psi$ (Assumption 2):
$$
\| \psi(\mathbf{PDD}^{(t)}) \| \leq \beta \| \mathbf{e}^{(t)} \| + \alpha \epsilon_{\text{PDD}}
$$
where $\alpha$ is a constant relating PDD error to gradient perturbation.

Step 5: Combine error terms.
Substituting into the error bound:
$$
\begin{aligned}
\|\mathbf{e}^{(t+1)}\| &\leq \rho(\mathbf{I} - \eta \mathbf{J}^{T}\mathbf{J}) \cdot \| \mathbf{e}^{(t)} \| + \eta \beta \| \mathbf{e}^{(t)} \| + \eta \alpha \epsilon_{\text{PDD}} \\
&= \left( \rho(\mathbf{I} - \eta \mathbf{J}^{T}\mathbf{J}) + \eta \beta \right) \| \mathbf{e}^{(t)} \| + \eta \alpha \epsilon_{\text{PDD}}
\end{aligned}
$$

Step 6: Enforce contraction mapping.
For linear convergence, we require:
$$
\rho(\mathbf{I} - \eta \mathbf{J}^{T}\mathbf{J}) + \eta \beta < 1
$$
Using the eigenvalue bound $\lambda_{\max}(\mathbf{J}^{T}\mathbf{J}) \leq J_{\max}^{2}$:
$$
\max_{i} |1 - \eta \lambda_{i}(\mathbf{J}^{T}\mathbf{J})| + \eta \beta < 1
$$
This yields the learning rate condition:
$$
\eta < \frac{2}{\lambda_{\max}(\mathbf{J}^{T}\mathbf{J}) + \beta}
$$

Step 7: Control PDD error perturbation.
From the PDD error bound (Assumption 4):
$$
\epsilon_{\text{PDD}} < \frac{\eta \gamma_{e}}{2 \| \mathbf{W}_{\text{att}} \|_{F}}
$$
This ensures that the perturbation term $\eta \alpha \epsilon_{\text{PDD}}$ remains dominated by the contraction term when:
$$
\eta \alpha \epsilon_{\text{PDD}} \ll \left( \rho(\mathbf{I} - \eta \mathbf{J}^{T}\mathbf{J}) + \eta \beta \right) \| \mathbf{e}^{(t)} \|
$$
which holds under the specified bound.

Step 8: Verify convergence.
Under both conditions, the error decreases geometrically:
$$
\|\mathbf{e}^{(t+1)}\| \leq \left( \rho(\mathbf{I} - \eta \mathbf{J}^{T}\mathbf{J}) + \eta \beta \right) \| \mathbf{e}^{(t)} \| + \eta \alpha \epsilon_{\text{PDD}} < \| \mathbf{e}^{(t)} \|
$$

\printbibliography 
\end{document}